\documentclass[twocolumn,
amsmath,amssymb,aps,prb,footinbib,floatfix,%
balancelastpage,
superscriptaddress,10pt]{revtex4-1}
\pdfoutput=1 

\makeatletter\providecommand{\galley@sw}[2]{}\galley@sw{%
\usepackage[notref,notcite]{showkeys}
\AtBeginDocument{\addtolength{\hoffset}{0.4\columnwidth}}%
}{}\preprint@sw{\def\Dated@name{\protect\texttt{\protect\jobname.tex~}-~}%
\def\@evenhead{\thepage\hfil{\tiny\@date}\hfil\hphantom{\thepage}\slshape\leftmark}%
\def\@oddhead{{\slshape\rightmark}\hphantom{\thepage}\hfil{\tiny\@date}\hfil\thepage}%
}{
}\makeatother%

\usepackage[utf8]{inputenc}

\usepackage{graphicx}
\usepackage{mathtools}
\usepackage{hyperref}
\usepackage{bm,bbm}
\usepackage{amsmath,amssymb,amsfonts}
\usepackage{color}

\newcommand{\DX}{\ensuremath{\text{d}x}}
\newcommand{\DY}{\ensuremath{\text{d}y}}
\newcommand{\DR}{\ensuremath{\text{d}r}}

\newcommand{\IMI}{{\text{i}}}
\newcommand{\EXP}{{\text{e}}}

\newcommand{\REXPO}{c}

\newcommand{\expval}[1]{\langle{#1}\rangle}

\newcommand{\oket}[1]{|{#1}\rangle_0}

\newcommand{\mydag}{+}
\newcommand{\phdag}{{\vphantom{\mydag}}}
 
\newcommand{\PRIME}{^{\prime\!}}

\newcommand{\AST}{{}^\ast_\ast}
\newcommand{\ORD}[1]{\AST{#1}\AST}
\newcommand{\Cadd}[1]{c^{\mydag}_{#1}}
\newcommand{\Cdel}[1]{c^{\phdag}_{#1}}
\newcommand{\Badd}[1]{b^{\mydag}_{#1}}
\newcommand{\Bdel}[1]{b^{\phdag}_{#1}}
\newcommand{\BIGBadd}[1]{B^{\mydag}_{#1}}
\newcommand{\BIGBdel}[1]{B^{\phdag}_{#1}}
\newcommand{\LBadd}[1]{B^{\mydag}_{#1}}
\newcommand{\LBdel}[1]{B^{\phdag}_{#1}}
\newcommand{\PSIadd}[2]{\psi_{#1}^\mydag(#2)}
\newcommand{\PSIdel}[2]{\psi_{#1}^\phdag(#2)}
\newcommand{\XIadd}[1]{\Xi_{#1}^\mydag}
\newcommand{\XIdel}[1]{\Xi_{#1}^\phdag}

\newcommand{\PHIadd}[2]{\varphi_{#1}^{\mydag\!}(#2)}
\newcommand{\PHIdel}[2]{\varphi_{#1}^{\phdag}(#2)}

\newcommand{\Fadd}[1]{F_{#1}^\mydag}
\newcommand{\Fdel}[1]{F_{#1}^\phdag}

\newcommand{\calFdel}[1]{{\cal F}_{#1}^\phdag}
\newcommand{\K}{K^{\vphantom{(}}}

\newcommand{\hatN}{\hat{N}}

\newcommand{\hatH}{\hat{H}}

\newcommand{\calC}{{\cal C}}
\newcommand{\calJ}{{\cal J}}
\newcommand{\calN}{{\cal N}}
\newcommand{\calhatN}{\hat{\cal N}}
\newcommand{\calhatJ}{\hat{\cal J}}

\newcommand{\BELL}{\mathfrak{B}}

\newcommand{\EMPTY}{{\eta}}
\newcommand{\EMPTYPRIME}{{\eta'}}
\newcommand{\SPACEEMPTY}{{,\eta}}
\newcommand{\NODELTA}{{\delta}}
\newcommand{\NOETA}{{}}

\newcommand{\VF}{v_{\text{F}}}
\newcommand{\Ggtr}{G^>}
\newcommand{\Gles}{G^<}
\newcommand{\Ggtrles}{G^{\gtrless}}
\newcommand{\ch}{\cosh}
\newcommand{\sh}{\sinh}
\newcommand{\tnh}{\mathop{\text{tanh}}}
\newcommand{\cth}{\mathop{\text{coth}}}
\newcommand{\COSHTHETA}{{u}}
\newcommand{\SINHTHETA}{{v}}

\begin{document}

  \title{From~Luttinger~liquids~to~Luttinger~droplets via~higher-order~bosonization~identities}

  \author{Sebastian Huber}

  \affiliation{Theoretical Solid State Physics, Arnold Sommerfeld
    Center for Theoretical Physics, Center for NanoScience, and Munich
    Center for Quantum Science and Technology,
    Ludwig-Maximilians-University, Theresienstr. 37, 80333 Munich,
    Germany}

  \author{Marcus Kollar}

  \affiliation{Theoretical Physics III, Center for Electronic
    Correlations and Magnetism, Institute of Physics, University of
    Augsburg, 86135 Augsburg, Germany}

  \date{\today}
  
  \begin{abstract}
    We derive generalized Kronig identities expressing quadratic
    fermionic terms including momentum transfer to bosonic operators
    and use them to obtain the exact solution for one-dimensional
    fermionic models with linear dispersion in the presence of
    position-dependent interactions and scattering potential. In these
    Luttinger droplets, which correspond to Luttinger liquids with
    spatial variations or constraints, the position dependences of the
    couplings break the translational invariance of correlation
    functions and modify the Luttinger-liquid interrelations between
    excitation velocities.
  \end{abstract}

  
  \maketitle

  \section{Introduction}\label{sec:introduction}

  An important goal of condensed matter theory is a reliable
  description of the correlated behavior of electrons which is rooted
  in the
  Coulomb interaction between them. In one-dimensional geometries they
  exhibit a special coherence at low
  energies:\cite{senechal_introduction_1999,giamarchi_quantum_2003} 
  the dispersion can be approximately linearized in the vicinity of
  the Fermi points $\pm k_{\text{F}}$ as $\epsilon_k$ $\simeq$
  $\pm \VF(k\pm k_\text{F})$, so that the energy $\delta\epsilon$ $=$
  $\VF\delta k$ of a particle-hole excitation from $k_1$ to
  $k_2$ is a function only of the momentum transfer $\delta k$ $=$
  $k_2-k_1$. By contrast, in higher dimensions the magnitude and
  relative orientation of the two momenta usually enter into
  $\delta\epsilon$, leading to a continuum of excitation energies for
  a given momentum transfer. This coherence in one dimension is
  prominently featured in the Tomonaga-Luttinger
  model,\cite{tomonaga_remarks_1950,luttinger_exactly_1963} which is
  based on the approximation that one can regard a physical electron
  field operator $\Psi(x)$ for a wire of length $L$ at low energies as
  a sum of two independent fields,
  \begin{subequations}\label{eq:psiphys}%
    \begin{align}
      \Psi(x)  
      &=
        \sum_{k}
        \frac{e^{\IMI kx}}{\sqrt{L}}
        \calC\phdag_k
      =
      \sum_{\substack{
      k>-k_\text{F}\\
      \lambda=\pm
      }}
      \frac{e^{\IMI\lambda(k_\text{F}+k)x}}{\sqrt{L}}
      \calC\phdag_{\lambda(k_\text{F}+k)}
      \label{eq:psiphysdef}
      \\&
      \simeq
      \frac{
      e^{\IMI k_\text{F}x}
      \PSIdel{\text{R}}{x}
      +
      e^{-\IMI k_\text{F}x}
      \PSIdel{\text{L}}{x}
      }{\sqrt{2\pi}}
      \label{eq:psiphysapprox}
      \,,
    \end{align}%
  \end{subequations}
  \begin{figure}[t]
    \centering
    \includegraphics[width=\columnwidth]{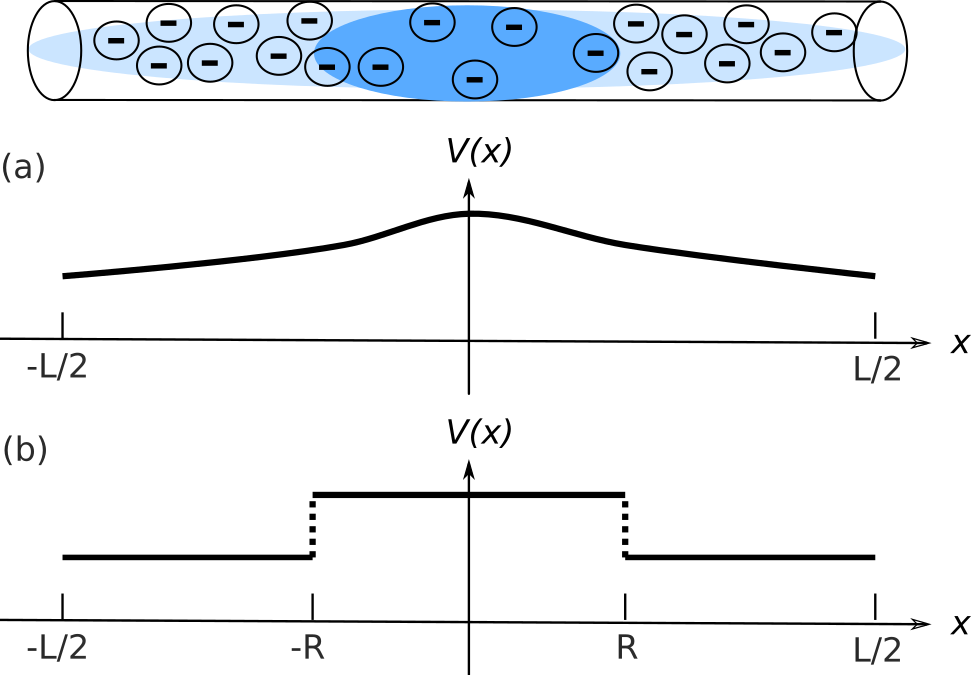}
    \caption{Wire of length $L$ with position-dependent interaction
      potential $V(x)$ in~\eqref{eq:dropletHfermionicLR-intro}, with
      $V(x)$ $=$ $V(-x)$, sketched here for the repulsive case with
      larger $V(x)$ near $x$ $=$ $0$ so that particles tend to keep a
      larger distance from each other there. An additional
      single-particle potential $W(x)$ $=$ $W(-x)$ may also be
      present. (a) A general smooth interaction potential. (b) A
      piecewise constant interaction potential, i.e., with piecewise
      constant value $V(0)$ inside and $V(L/2)$ outside a central
      region of width $2R$, as solved explicitly in
      Sec.~\ref{subsubsec:piecewise} for $L$ $\to$ $\infty$ and finite
      $R$.\label{fig:sketch}}
  \end{figure}
  where the lower summations limits $-k_{\text{F}}$
  in~\eqref{eq:psiphysdef} were replaced by
  $-\infty$~\eqref{eq:psiphysapprox}. Then $\PSIdel{\text{R,L}}{x}$
  $=$ $\PSIdel{1,2}{\mp x}$ and $\PSIdel{\eta}{x}$ $=$
  $(2\pi/L)^{\frac12}$ $\sum_{k} \EXP^{-\IMI kx} \Cdel{k\eta}$
  are defined in terms of canonical fermions $\Cdel{k\eta}$ $=$
  $\calC\phdag_{\pm(k_\text{F}+k)}$ which correspond to the
  physical fermions $\calC_k$ near the two Fermi points for $\eta$
  $=$ $1,2$.  In the Tomonaga-Luttinger model the dispersion is
  linearized near the Fermi points and only forward-scattering density
  interactions between left- and right-moving fermions are kept. The
  Tomonaga-Luttinger model can be solved by
  bosonization,\cite{tomonaga_remarks_1950,luttinger_exactly_1963,mattis_exact_1965,%
    schick_flux_1968,%
    schotte_tomonagas_1969,%
    mattis_new_1974,%
    luther_single-particle_1974,%
    coleman_quantum_1975,mandelstam_soliton_1975,%
    heidenreich_sine-gordon_1975,%
    haldane_coupling_1979,%
    *haldane_luttinger_1981} which expresses the above-mentioned
  coherence of excitations into an exact mapping to bosonic degrees of
  freedom (at the operator level\cite{luther_single-particle_1974,%
    emery_theory_1979,%
    voit_one-dimensional_1995,%
    kotliar_toulouse_1996,%
    von_delft_bosonization_1998,%
    von_delft_finite-size_1998,%
    *zarand_analytical_2000} or in a path-integral
  formulation;\cite{lee_functional_1988,%
    yurkevich_bosonisation_2002,%
    grishin_functional_2004,%
    galda_impurity_2011,%
    filippone_tunneling_2016} throughout we use
  Ref.~\onlinecite{von_delft_bosonization_1998}'s constructive
  finite-size bosonization approach, which is recapped
  below). Bosonization has led to such remarkable results and concepts
  as spin-charge separation of elementary
  excitations,\cite{tomonaga_remarks_1950,luttinger_exactly_1963}
  interaction-dependent exponents of correlation
  functions,\cite{luther_single-particle_1974,%
    meden_spectral_1992,%
    *schonhammer_nonuniversal_1993,%
    *schonhammer_erratum:_1993,%
    meden_nonuniversality_1999,%
    markhof_spectral_2016} and the Luttinger-liquid
  paradigm\cite{haldane_luttinger_1981,%
    haldane_general_1980,%
    *haldane_demonstration_1981,%
    *haldane_effective_1981} which states that the relations between
  excitation velocities and correlation exponents of the
  Tomonaga-Luttinger model remain valid even for weakly nonlinear
  dispersion. These topics are nowadays presented in many
  reviews\cite{senechal_introduction_1999,%
    emery_theory_1979,%
    solyom_fermi_1979,%
    voit_one-dimensional_1995,%
    schonhammer_interacting_1997,%
    *schonhammer_luttinger_2004,%
    *schonhammer_physics_2013,%
    von_delft_bosonization_1998,%
    miranda_introduction_2003,%
    cazalilla_one_2011} and
  textbooks.\cite{kopietz_bosonization_1997,%
    giamarchi_quantum_2003,%
    gogolin_bosonization_2004,%
    bruus_many-body_2004,%
    giuliani_quantum_2008,%
    phillips_advanced_2012,%
    mastropietro_luttinger_2013} Characteristic signatures of
  one-dimensional electron liquids have been observed in a variety of
  experiments.\cite{%
    milliken_indications_1996,%
    maasilta_line_1997,%
    chang_chiral_2003,%
    bockrath_luttinger-liquid_1999,%
    ishii_direct_2003,%
    aleshin_one-dimensional_2004,%
    boninsegni_luttinger_2007,%
    *del_maestro_4he_2011,%
    *duc_critical_2015,%
    jompol_probing_2009,%
    barak_interacting_2010,%
    blumenstein_atomically_2011,%
    mebrahtu_quantum_2012,%
    *mebrahtu_observation_2013,%
    yang_quantum_2017,%
    cedergren_insulating_2017,%
    stuhler_tomonaga-luttinger_2019%
  }
  The theory of nonlinear dispersion terms has been
  of particular further interest,\cite{schick_flux_1968,%
    haldane_luttinger_1981,%
    busche_how_2000,%
    *pirooznia_dynamic_2008,%
    teber_bosonization_2007,%
    karrasch_low_2015} including refermionization techniques which use
  bosonization identities in reverse to map diagonalized bosonic
  systems back to free fermions.\cite{rozhkov_variational_2003,%
    *rozhkov_fermionic_2005,%
    *rozhkov_class_2006,%
    *rozhkov_density-density_2008,%
    *rozhkov_one-dimensional_2014,%
    imambekov_universal_2009,%
    *imambekov_phenomenology_2009,%
    *imambekov_one-dimensional_2012,%
    teber_bosonization_2007,%
    maebashi_structural_2014,%
    essler_spin-charge-separated_2015,%
    markhof_investigating_2019}
  For Luttinger liquids out of
  equilibrium\cite{cazalilla_effect_2006,%
    *iucci_quantum_2009,%
    *nessi_quantum_2013,%
    uhrig_interaction_2009,%
    foster_quantum_2010,%
    perfetto_thermalization_2011,%
    dziarmaga_excitation_2011,%
    dora_crossover_2011,%
    *dora_generalized_2012,%
    *dora_absence_2015,%
    *dora_momentum-space_2016,%
    karrasch_luttinger-liquid_2012,%
    *rentrop_quench_2012,%
    coira_quantum_2013,%
    ngo_dinh_interaction_2013,%
    sabetta_nonequilibrium_2013,%
    kennes_luttinger_2013,%
    *kennes_spectral_2014,%
    sachdeva_finite-time_2014,%
    schiro_transport_2015,%
    mastropietro_quantum_2015} nonlinear dispersion effects are also
  essential.\cite{gutman_bosonization_2010,%
    *protopopov_many-particle_2011,
    *protopopov_relaxation_2014,%
    *protopopov_dissipationless_2014,%
    *protopopov_equilibration_2015,%
    lin_thermalization_2013,%
    buchhold_nonequilibrium_2015,%
    *buchhold_kinetic_2015,%
    *buchhold_prethermalization_2016,%
    *huber_thermalization_2018}%

  
  The technical hallmarks of bosonization are the following. On the
  one hand, a two-body density interaction term for fermions
  $\Cdel{k\EMPTY}$ becomes quadratic in terms of canonical bosons,
  defined for $q$ $>$ $0$ as $\Bdel{q\EMPTY}$ $=$
  $-\IMI\sum_k\Cadd{k-q\EMPTY}\Cdel{k\EMPTY}/\sqrt{\smash[b]{n_q}}$. Here
  the momentum sum runs over $k$ $=$
  $\frac{2\pi}{L} (n_k-\tfrac12\delta_\text{b})$ with integer $n_k$,
  and the parameter $0$ $\leq$ $\delta_\text{b}$ $<$ $2$ fixes the
  boundary conditions, $\PSIdel{\EMPTY}{x+L/2}$ $=$
  $\EXP^{i\pi\delta_\text{b}}$ $\PSIdel{\EMPTY}{x-L/2}$.  On the
  other hand the fermionic kinetic energy also translates into free
  bosons by means of the so-called Kronig
  identity,\cite{kronig_zur_1935,dover_properties_1968,schonhammer_interacting_1997}
  \begin{subequations}\label{eq:H01}%
    \begin{align}%
      H_{0\EMPTY}^{(1)}
      &=
        \sum_k
        k
        \,
        \ORD{
        \Cadd{k\EMPTY}
        \Cdel{k\EMPTY}
        }
        \label{eq:H01-fermionic}
      \\&
      =   
      \sum_{q>0}
      q
      \,
      \Badd{q\EMPTY}
      \Bdel{q\EMPTY}
      +
      \frac{\pi}{L}
      (\hatN_{\EMPTY}+1-\delta_\text{b})\hatN_{\EMPTY}
      \,,
      \label{eq:H01-bosonic}
    \end{align}\end{subequations}
  where the fermionic number operator is given by\footnote{Throughout,
    hats appear only on those operators which involve fermionic number
    operators.} $\hatN_{\EMPTY}$ $=$
  $\sum_k\ORD{\Cadd{k\EMPTY}\Cdel{k\EMPTY}}$, which commutes with
  $\Bdel{q\EMPTY}$.  The normal ordering $\ORD{\cdots}$ is defined with
  respect to the state $\oket{\bm{0}}$, where $\oket{\bm{N}}$ is an
  eigenstate of all $\Cadd{k\EMPTY}\Cdel{k\EMPTY}$ (with eigenvalue 1 if
  $n_k$ $\leq$ $N_\EMPTY$ and 0 otherwise).  Furthermore, real-space
  fermionic and bosonic fields are related by the celebrated
  bosonization identity,\cite{schotte_tomonagas_1969,%
    mattis_new_1974,%
    luther_single-particle_1974,%
    von_delft_bosonization_1998}
  \begin{align}
    \PSIdel{\EMPTY}{x}
    &=
      \big(\tfrac{2\pi}{L}\big)^{\frac12}
      \Fdel{\EMPTY}
      \,
      \EXP^{-\IMI\frac{2\pi}{L}(\hatN_{\EMPTY}-\frac12\delta_\text{b})x}
      \,
      \EXP^{-\IMI\PHIadd{\EMPTY}{x}}
      \,
      \EXP^{-\IMI\PHIdel{\EMPTY}{x}}
      ,
      \label{eq:intro:bosonizationidentity}
  \end{align}
  which allows the calculation of fermionic in terms of bosonic
  correlation functions.\cite{luther_single-particle_1974} Here the fermionic
  Klein factor $\Fdel{\EMPTY}$ decreases the fer\-mi\-on\-ic particle
  number $\hatN_{\EMPTY}$ by one,
  and
  $\PHIdel{\EMPTY}{x}$ $=$
  $-\sum_{q>0} \Bdel{q\EMPTY}\,\EXP^{-\IMI
    qx-aq}/\sqrt{\smash[b]{n_q}}$. The regularization parameter $a$
  $\to$ $0^+$ is needed to obtain a finite
  commutator,\cite{von_delft_bosonization_1998}
  $[\PHIdel{\EMPTY}{x},\PHIadd{\EMPTY}{x\PRIME}]$ $=$
  $-\ln\!\big( 1-\EXP^{-\frac{2\pi\IMI}{L}(x-x\PRIME-\IMI a)}
  \big)$. A final ingredient to the solution of the Tomonaga-Luttinger
  model is a Boguljubov transformation, which absorbs the interaction
  between left- and right-moving fermions into the free bosonic
  theory.\cite{tomonaga_remarks_1950}


  In the present work we will study Luttinger liquids with additional
  spatial constraints, which we term \emph{Luttinger
    droplets}. Namely, we consider a (spinless) fermionic Hamiltonian with linear dispersion,
  position-dependent interactions $V(x)$ and $U(x)$, and scattering
  potential $W(x)$ (all assumed to be real symmetric functions of
  $x$),
  \begin{subequations}%
    \label{eq:dropletHfermionicLR-intro}%
    \begin{align}%
      \label{eq:dropletHfermionicLR-intro-Hamiltonian}
      H
      &=
        \int
        \!
        \frac{\DX}{L}\,
        \AST
        \VF[
        \PSIadd{\text{R}}{x}
        \IMI\partial_x
        \PSIdel{\text{R}}{x}
        -
        \PSIadd{\text{L}}{x}
        \IMI\partial_x
        \PSIdel{\text{L}}{x}
        ]
        \nonumber\\&\,\hphantom{=}\,
      +
      W(x)[n_{\text{L}}(x)+n_{\text{R}}(x)]
      +
      U(x)
      n_{\text{L}}(x)n_{\text{R}}(x)
      \nonumber\\&\,\hphantom{=}\,
      +
      \frac12V(x)[n_{\text{L}}(x)^2+n_{\text{R}}(x)^2]
      \AST
      \,,
    \end{align}
    in terms of densities $n_{\text{R,L}}(x)$ $=$
    $\PSIadd{\text{R,L}}{x}\PSIdel{\text{R,L}}{x}/(2\pi)$.  Without
    $W(x)$ and with constant $V(x)$ and $U(x)$, $H$ 
    reduces to the usual translationally invariant Tomonaga-Luttinger
    model (with contact interactions).  Below we will
    diagonalize~\eqref{eq:dropletHfermionicLR-intro-Hamiltonian} exactly for the
    special case
    \begin{align}
      U(x)
      &=
        \gamma\,[2\pi \VF+V(x)]
        \,,~~~~
        -1<\gamma<1
        \,,
        \label{eq:dropletHfermionicLR-intro-parameters}
    \end{align}      
  \end{subequations}
  for otherwise arbitrary $V(x)$ $>$
  $-2\pi\VF$ and a constant
  $\gamma$. This means that real Fourier components $V_q$ $=$
  $V_{-q}$ as well as $U_{q=0}$ can be chosen freely; then
  $\gamma$ $=$ $U_0/[2\pi \VF+V_0]$ and $U_{q\neq0}$ $=$ $\gamma
  V_q$. Thus
  $\gamma$ characterizes the relative strength of interbranch
  interactions.  Below we obtain the single-particle Green function
  for the ground state of this model, the exponents of which will
  reflect the spatial dependence of the couplings. We first derive
  generalized Kronig-type identities in Sec.~\ref{sec:kronig}, which
  we then use to solve a single-flavor chiral version
  of~\eqref{eq:dropletHfermionicLR-intro} in Sec.~\ref{sec:chiral}. We
  then proceed to the two-flavor case in Sec.~\ref{sec:droplet}, with
  a discussion of the similiarities and differences of the spectrum
  and Green function compared to the translationally invariant case.
  One representative choice of
  $V(x)$ to be discussed below involves a central region with stronger
  repulsion than at the edges of the system, as shown in
  Fig.~\ref{fig:sketch}a. An explicit evaluation is provided for a
  piecewise constant
  $V(x)$ as shown in Fig.~\ref{fig:sketch}b in
  Sec.~\ref{subsubsec:piecewise}.  Relations between the excitation
  velocities and Green function exponents are discussed in
  Sec.~\ref{sec:dropletparadigm}, followed by a summary in
  Sec.~\ref{sec:conclusion}.
  
  Many results for inhomogeneous Luttinger liquids are of course
  known, e.g., with barriers,\cite{%
    kane_transport_1992,%
    *kane_transmission_1992,%
    *kane_resonant_1992,%
    rylands_quantum_2016,%
    *rylands_quantum_2017,%
    *rylands_quantum_2018%
  } %
  impurities,\cite{meden_single_2002,hattori_quantum_2014} %
  boundaries,\cite{meden_luttinger_2000,schneider_recursive_2010,rylands_exact_2019} %
  leads,\cite{eggert_scanning_2000,filippone_tunneling_2016} %
  confinements\cite{wonneberger_luttinger-model_2001} %
  and so on. Models with (effective) position-dependent Luttinger
  liquid parameters or interaction potentials have also been
  investigated.\cite{%
    maslov_landauer_1995,%
    safi_transport_1995,%
    ponomarenko_renormalization_1995,%
    rech_electronic_2008,%
    *rech_resistivity_2008, grishin_functional_2004,
    *galda_impurity_2011} Our goal is to provide a complementary
  perspective on these setups with the exact solution of the rather
  flexible model~\eqref{eq:dropletHfermionicLR-intro}, i.e., the
  Hamiltonian~\eqref{eq:dropletHfermionicLR-intro-Hamiltonian} with
  parameters from the
  manifold~\eqref{eq:dropletHfermionicLR-intro-parameters}, and to
  possibly enable new applications, e.g., to ultradilute quantum
  droplets held together by weak cohesive
  forces.\cite{ferrier-barbut_ultradilute_2019}


  \section{Kronig-type identities with arbitrary momentum transfer}\label{sec:kronig}

  \subsection{Bosonic forms of bilinear fermionic terms}

  Consider a general bilinear fermionic term,
  \begin{subequations}\label{eq:intro:Hqm}%
    \begin{align}%
      \label{eq:intro:Hqm-k}
      H_{q\EMPTY}^{(m)}
      &=
        \sum_k
        k^m
        \,
        \ORD{
        \Cadd{k-q\EMPTY}
        \Cdel{k\EMPTY}
        }
      \\&
      \label{eq:intro:Hqm-x}
      =
      \int
      \!
      \frac{\DX}{2\pi}\,
      \EXP^{\IMI qx}
      \,
      \ORD{
      \PSIadd{\EMPTY}{x}
      (\IMI\partial_x)^m
      \PSIdel{\EMPTY}{x}
      }
      \,,
    \end{align}\end{subequations}
  for integer exponents $m$ $\geq$ $0$ and momentum transfer $q$ $=$
  $\frac{2\pi}{L}$ $n_q$ with integer $n_q$; here and throughout
  real-space integrals without indicated endpoints extend over the interval $[-L/2,L/2]$.
  Arbitrary
  dispersion terms are included in~\eqref{eq:intro:Hqm-k} for $q$ $=$
  $0$, such as~\eqref{eq:H01-fermionic} for $m$ $=$ $1$.
  Forming the product
  of~\eqref{eq:intro:bosonizationidentity} with its hermitian
  conjugate at different positions $x$ and $x+\ell$, canceling the
  Klein factors ($\Fadd{\EMPTY}\Fdel{\EMPTY}$ $=$ $1$), commuting the
  bosonic fields, taking $a$ to zero, and combining exponentials, we
  obtain
  \begin{align}
    &\frac{L}{2\pi}
      \,
      \PSIadd{\EMPTY}{x}
      \PSIdel{\EMPTY}{x+\ell}
      \label{eq:intro:psidagpsi-x}
    \\&
    =
    \EXP^{\pi\IMI(\delta_\text{b}-2\hatN_{\EMPTY})\ell/L}
    \,
    \EXP^{\IMI\PHIadd{\EMPTY}{x}}
    \,
    \EXP^{\IMI\PHIdel{\EMPTY}{x}}
    \,
    \EXP^{-\IMI\PHIadd{\EMPTY}{x+\ell}}
    \,
    \EXP^{-\IMI\PHIdel{\EMPTY}{x+\ell}}
    \nonumber\\&
    =
    \frac{
    \EXP^{\pi\IMI(\delta_\text{b}-2\hatN_{\EMPTY})\ell/L}
    }{1-\EXP^{2\pi\IMI\ell/L}}
    \,
    \EXP^{\IMI(\PHIadd{\EMPTY}{x}-\PHIadd{\EMPTY}{x+\ell})}
    \,
    \EXP^{\IMI(\PHIdel{\EMPTY}{x}-\PHIdel{\EMPTY}{x+\ell})}
    \,,\nonumber
  \end{align}
  A generating function of the terms in~\eqref{eq:intro:Hqm-k} then
  reads
  \begin{align}
    \label{eq:intro:Hq}
    &\sum_{m=0}^{\infty}
      \frac{(-\IMI\ell)^{m}}{m!}
      \,
      H_{q\EMPTY}^{(m)}
      =
      \sum_k
      \EXP^{-\IMI k\ell}
      \,
      \ORD{
      \Cadd{k-q\EMPTY}
      \Cdel{k\EMPTY}
      }
    \\&
    =
    \int
    \!\frac{\DX}{2\pi}\,
    \EXP^{\IMI qx}
    \,
    \ORD{
    \PSIadd{\EMPTY}{x}
    \PSIdel{\EMPTY}{x+\ell}
    }
    =
    \int
    \!\frac{\DX}{L}\,
    \frac{
    \EXP^{\pi\IMI\delta_\text{b}\ell/L}
    \,
    \EXP^{\IMI qx}
    }{1-\EXP^{2\pi\IMI\ell/L}}
    \nonumber\\&~~~\times
    \,
    \big(
    \EXP^{-2\pi\IMI\hatN_{\EMPTY}\ell/L}
    \,
    \EXP^{\IMI\PHIadd{\EMPTY}{x}-\IMI\PHIadd{\EMPTY}{x+\ell}}
    \,
    \EXP^{\IMI\PHIdel{\EMPTY}{x}-\IMI\PHIdel{\EMPTY}{x+\ell}}-1\big)
    \,,\nonumber
  \end{align}
  where we summed the Taylor series of the
  terms~\eqref{eq:intro:Hqm-x}, inserted
  relation~\eqref{eq:intro:psidagpsi-x}, and performed the normal
  ordering. Taylor expanding the exponentials and taking coefficients
  of $\ell^m$ on both sides of~\eqref{eq:intro:Hq} now yields
  $H_{q\EMPTY}^{(m)}$ in terms of bosonic operators, as discussed
  below.  Relation~\eqref{eq:intro:Hq} thus provides explicit bosonic
  representations of general bilinear fermionic operators,
  including~\eqref{eq:H01}.\footnote{We note that our derivation
    of~\eqref{eq:intro:Hq} is similar in spirit to the procedure in
    Sec.~3.2 of Ref.~\onlinecite{pereira_dynamical_2007}, and also
    bears some resemblance to the analysis of higher-order dispersion
    terms in
    Ref.~\onlinecite{enciso_fermion_2006,*karabali_exact_2014}. However
    our approach is more general since we also allow finite momentum
    transfer $q$ and work with exact operator identities. Note also
    that one may view~\eqref{eq:intro:Hq} as a fermionic
    representation of certain properties of vertex
    operators,\cite{von_delft_bosonization_1998} i.e., exponentiated
    bosonic fields.}%
  \nocite{pereira_dynamical_2007,enciso_fermion_2006,karabali_exact_2014,von_delft_bosonization_1998}
  %


  We also introduce operators which use the more convenient powers of
  the integer $n_k$ instead of momentum $k$,
  \begin{subequations}\label{eq:K-def}%
    \begin{align}%
      \label{eq:Kq}
      \!\!\!\!\!\!
      \!\!\!\!\!\!
      \!\!\!\!\!\!
      \!\!\!\!\!\!
      \K_{q\EMPTY}(\lambda)
      &=
        \sum_k
        \EXP^{\lambda n_k}
        \;
        \ORD{
        \Cadd{k-q\EMPTY}
        \Cdel{k\EMPTY}
        }
        =
        \sum_{m=0}^\infty
        \frac{\lambda^m}{m!}
        \,
        K_{q\EMPTY}^{(m)}
        \,,    
      \\
      \label{eq:Kqm}
      K_{q\EMPTY}^{(m)}
      &=
        \sum_k
        n_{k}^{m}
        \;
        \ORD{
        \Cadd{k-q\EMPTY}
        \Cdel{k\EMPTY}
        }
        \,,
      \\
      \label{eq:intro:Kq0}
      K_{q\EMPTY}^{(0)}
      &=
        \sum_k
        \ORD{
        \Cadd{k-q\EMPTY}
        \Cdel{k\EMPTY}
        }
      =
      \begin{cases}
        \phantom{-}\hatN_{\EMPTY}
        &
        \text{\!\!\!\!if\,}
        n_q\!=\!0,
        \\
        \phantom{-}\IMI\sqrt{\smash[b]{n_q}}\phantom{{}_{-}}\,\Bdel{q\EMPTY}
        &\text{\!\!\!\!if\,}
        n_q\!>\!0,
        \\
        -\IMI\sqrt{n_{-q}}\,\Badd{-q\EMPTY}
        &\text{\!\!\!\!if\,}
        n_{q}\!<\!0,
      \end{cases}
          \!\!\!\!\!\!
    \end{align}\end{subequations}
  so that the terms~\eqref{eq:intro:Hqm-k} are then given by $H_{q\EMPTY}^{(m)}$
  $=$ $(\pi/L)^m$ $\sum_{n=0}^{m}$ $\binom{m}{n}$
  $(-\delta_\text{b})^{m-n}$ $2^n$ $K_{q\EMPTY}^{(n)}$ and the bosonic commutation relations become
  $[K_{-q\EMPTY}^{(0)},K_{q\PRIME\EMPTYPRIME}^{(0)}]$ $=$ $-n_q\delta_{qq\PRIME}\delta_{\EMPTY\EMPTYPRIME}$.
  The
  operators $\K_{q\EMPTY}(\lambda)$, which are 
  operator-valued formal power series in the (complex) indeterminate
  $\lambda$ with coefficients $K_{q\EMPTY}^{(m)}$,  obey the intriguing operator algebra
  \begin{align}
    \label{eq:intro:comm1}
    \!\!\Big[\K_{-q\EMPTY}(\lambda),\K_{q\PRIME\EMPTYPRIME}(\lambda\PRIME)\Big]
    &=
      \NODELTA_{\EMPTY\EMPTYPRIME}\bigg[
      \delta_{qq\PRIME}\frac{\EXP^{-\lambda n_{q}}-\EXP^{\lambda\PRIME n_{q}}}{1-\EXP^{-\lambda-\lambda\PRIME}}
    \\&~~~
    +
    (\EXP^{-\lambda n_{q\PRIME}}-\EXP^{\lambda\PRIME n_{q}})\K_{q\PRIME-q\EMPTY}(\lambda+\lambda\PRIME)
    \bigg],\nonumber
  \end{align}
  which is reminiscient of affine Lie algebras,\cite{francesco_conformal_1997} but not immediately recognizable.
  From~\eqref{eq:intro:Hq}, or alternatively
  from~\eqref{eq:intro:comm1}, the generating function~\eqref{eq:Kq}
  becomes
  \begin{subequations}\label{eq:intro:K-lambda-result}%
    \begin{align}%
      \!\K_{q\EMPTY}(\lambda)
      &=
        \frac{
        \EXP^{\lambda\hatN_{\EMPTY}}
        Y_{q\EMPTY}(\lambda)
        -
        \delta_{q0}
        }{1-\EXP^{-\lambda}}
        \,,
        \label{eq:intro:Kq-lambda-result}
      \\
      \!Y_{q\EMPTY}(\lambda)
      &=
        \sum_{n,r=0}^{\infty}
        \frac{1}{n!r!}
        \sum_{\substack{p_1,\ldots,p_n>0\\p_1\PRIME,\ldots,p_r\PRIME>0}}
      \delta_{p_1+\cdots p_n+q,p_1\PRIME+\cdots p_r\PRIME}
      \nonumber
      \\
      \nonumber
      &\;\times\!
        \Big(
        \prod_{i=1}^n
        \frac{1-\EXP^{-\lambda n_{p_i}}}{n_{p_i}}
        K_{-p_i\EMPTY}^{(0)}      
        \Big)
        \!
      \Big(
      \prod_{j=1}^r
      \frac{\EXP^{\lambda n_{p_j\PRIME}}-1}{n_{p_j\PRIME}}
      K_{p_j\PRIME\EMPTY}^{(0)}
      \Big)
      \\
      &=
      \sum_{m=0}^\infty
      \frac{\lambda^m}{m!}
      Y_{q\EMPTY}^{(m)}
      \,.
    \end{align}\end{subequations}
  The coefficients $K_{q\EMPTY}^{(m)}$ and $Y_{q\EMPTY}^{(m)}$ of
  $\lambda^m$ in these expression are given by
  \begin{subequations}\label{eq:intro:Kqm:all}%
    \begin{align}%
      K_{q\EMPTY}^{(m)}
      &=
        \sum_{m=0}^\infty
        \frac{\lambda^m}{m!}
        \,
        K_{q\EMPTY}^{(m)}
        =
        \frac{
        B_{m+1}(\hatN_{\EMPTY}+1)-B_{m+1}(1)
        }{m+1}
        \delta_{q0}
        \nonumber
      \\&\label{eq:intro:Kqm}
      +
      \sum_{n=0}^{m}
      \binom{m}{n}
      \frac{
      (-1)^n
      }{
      m+1-n
      }
      B_n(-\hatN_{\EMPTY})
      \,
      Y_{q\EMPTY}^{(m+1-n)}
      \,,
      \end{align}
      \begin{align}
      Y_{q\EMPTY}^{(m)}
      &=
        \int
        \!\frac{\DX}{L}
        \,
        \EXP^{\IMI qx}
        \sum_{n=0}^{m}
        \binom{m}{n}
        \BELL_{m}\big(K_{+\SPACEEMPTY}^{(1)\!}(x),...,K_{+\SPACEEMPTY}^{(m)\!}(x)\big)
        \nonumber\\&\times
      \BELL_{m}\big(K_{-\SPACEEMPTY}^{(1)\!}(x),...,K_{-\SPACEEMPTY}^{(m)\!}(x)\big)
      \,,
    \end{align}\end{subequations}
  Here $K_{\pm\SPACEEMPTY}^{(m)\!}(x)$ $=$
  $\sum_{\pm p>0} n_{-p}^{m-1} K_{-p\EMPTY}^{(0)} \EXP^{\IMI px}$ and
  $B_n(x)$ and $\BELL_m(x_1,\ldots,x_m)$ are the Bernoulli and
  complete Bell polynomials, respectively, defined
  by\cite{comtet_advanced_1974}
  \begin{subequations}%
    \begin{align}%
      \frac{
      \lambda\,\EXP^{\lambda x}
      }{
      \EXP^{\lambda}-1
      }
      &=
        \sum_{m=0}^\infty
        \frac{\lambda^m}{m!}B_m(x)
        \,,\\
      \exp\!\bigg(
      \sum_{m=1}^\infty
      \frac{\lambda^m}{m!}\,x_m
      \bigg)
      &=
        \sum_{m=0}^\infty
        \frac{\lambda^m}{m!}
        \BELL_m(x_1,\ldots,x_m)
        \,.
    \end{align}\end{subequations}%
  A detailed derivation
  of~\eqref{eq:intro:K-lambda-result}-\eqref{eq:intro:Kqm:all} will be
  presented elsewhere.

  \subsection{Bosonic representation of a fermionic scattering term}

  Generalized Kronig identities for arbitrary order $m$ follow from
  the equivalence of~\eqref{eq:Kqm} and~\eqref{eq:intro:Kqm}, with the
  latter involving only fermionic number operators and normal-ordered
  bosonic operators.
  As a special case, we obtain for $m$ $=$ $1$ and $q$ $\neq$ $0$ the
  finite-$q$ generalization of~\eqref{eq:H01},
  \begin{align}
    \!\!\!
    K_{q\EMPTY}^{(1)}
    &=
      \sum_k
      n_k
      \,
      \ORD{
      \Cadd{k-q\EMPTY}
      \Cdel{k\EMPTY}
      }
      \nonumber\\&
    =
    \Big(
    \frac{n_q+1}{2}
    +
    \hatN_{\EMPTY}
    \Big)
    K_{q\EMPTY}^{(0)}
    +
    \frac12
    \sum_{p(\neq0,q)}\!\!
    K_{q-p\EMPTY}^{(0)}
    K_{p\EMPTY}^{(0)}
    \,,
    \label{eq:intro:Kq1-bosonic}
  \end{align}
  which can also be expressed as
  \begin{subequations}\label{eq:Hq1}%
    \begin{align}
      H_{q\EMPTY}^{(1)}
      &=
        \sum_k
        k
        \,
        \ORD{
        \Cadd{k-q\EMPTY}
        \Cdel{k\EMPTY}
        }
        =\frac{2\pi}{L}K_{q\EMPTY}^{(1)}-\frac{\pi\delta_{\text{b}}}{L}\hatN_{\EMPTY}
        \label{eq:Hq1-fermionic}
      \\&
      =
      \Big(\frac{q}{2}+\frac{\pi}{L}(2\hatN_{\EMPTY}+1-\delta_{\text{b}})\Big)
      \,
      \IMI\sqrt{\smash[b]{n_q}}\,
      \Bdel{q\EMPTY}
      \nonumber\\&~~~~
      -
      \frac12\!
      \sum_{q>p>0}
      \!
      \sqrt{(q-p)p}\,
      \Bdel{q-p\EMPTY}
      \Bdel{p\EMPTY}
      \nonumber\\&~~~~
      +
      \sum_{p>0}
      \sqrt{p(q+p)}\,
      \Badd{p\EMPTY}
      \Bdel{q+p\EMPTY}
      \,,
      ~~~~
      (q>0)
      \label{eq:Hq1-bosonic}
    \end{align}%
  \end{subequations}
  so as to make the modification of the momentum-diagonal
  identity~\eqref{eq:H01} more apparent.

  \section{Chiral Luttinger droplets}\label{sec:chiral}

  \subsection{Droplet model with only right movers}

  As a simple application of~\eqref{eq:Hq1} and for later reference we
  first consider a single species of spinless fermions with density
  \begin{align}
    n_{\NOETA}(x) =
    \frac{1}{2\pi}\,\PSIadd{\NOETA}{x}\PSIdel{\NOETA}{x} =
    \frac{1}{L}\,\sum_qK_{q\NOETA}^{(0)}e^{-iqx}\,,
  \end{align}
  subjected to a single-particle potential $w(x)$ $=$ $w(-x)$
  and a position-dependent interaction $g(x)$ $=$ $g(-x)$, with
  Fourier transforms $w_q$ $=$ $\int\!w(x)\,\EXP^{-\IMI qx}\DX/L$ $=$
  $w_{-q}$ and so on. For simplicity we choose antiperiodic boundary
  conditions ($\delta_{\text{b}}$ $=$ $1$). For a linear dispersion
  the Hamiltonian of such a `chiral Luttinger droplet' is given by
  %
  \begin{align}\label{eq:chiralH}
    H_{\text{chiral}}
    &=
      \VF
      \sum_{k}
      k
      \,
      \ORD{
      \Cadd{k\NOETA}
      \Cdel{k\NOETA}
      }
      +
      \int
      \!\frac{\DX}{L}\,
      w(x)\,
      \ORD{
      n_{\NOETA}(x) 
      }
    \\\nonumber&\,\hphantom{=}\,
                 +
                 \frac12
                 \int
                 \!
                 \frac{\DX}{L}\,
                 g(x)\,
                 \ORD{
                 n_{\NOETA}(x)^2      
                 }
                 \,.
  \end{align}

  \subsection{Diagonalization of the chiral model}

  On the one hand, we can now express the fermionic Hamiltonian
  $H_{\text{chiral}}$ in terms of bosonic operators. We define
  \begin{align}
    H_{[\bm{\tilde{g}},\hat{\bm{\tilde{w}}};\bm{K}]}^{\text{bosonic}}
    &=
      \frac{\tilde{g}_0}{L}
      \sum_{q>0}
      K_{-q\NOETA}^{(0)}
      K_{q\NOETA}^{(0)}
      \label{eq:chiralHbosonic}%
    \\&\,\hphantom{=}\,
    +
    \frac{1}{L}
    \sum_{q\neq0}
    \bigg[
    \hat{\tilde{w}}_q
    K_{q\NOETA}^{(0)}
    +
    \frac{\tilde{g}_q}{2}
    \!\sum_{p(\neq0,q)}\!
    K_{p\NOETA}^{(0)}
    K_{q-p\NOETA}^{(0)}
    \bigg]
    \,,
    \nonumber
  \end{align}%
  with symmetric parameters
  $\hat{\tilde{w}}_q$ (that may contain $\hatN$) and $\tilde{g}_q$. For $\tilde{g}_q$ $=$
  $2\pi \VF\delta_{q0}$ $+$ $g_q$ and $\hat{\tilde{w}}_q$ $=$ $w_qL$ $+$
  $g_q\hatN_{\NOETA}$ 
  we find that
  \begin{align}
    H_{\text{chiral}}
    &=
      H_{[\bm{\tilde{g}},\hat{\bm{\tilde{w}}};\bm{K}]}^{\text{bosonic}}
      +
      \frac{\tilde{g}_0}{2L}\hatN_{\NOETA}^2
      +
      w_0\hatN_{\NOETA}
      \,.
  \end{align}
  On the other hand, the fermionic basis permits a full
  diagonalization as follows. Using~\eqref{eq:intro:Kq1-bosonic} to
  eliminate the last term in~\eqref{eq:chiralHbosonic} we arrive at a
  fermionic scattering Hamiltonian,
  %
  \begin{align}\label{eq:chiralHfermionic}%
    H_{\text{chiral}}
    &=
      \sum_{kk\PRIME}
      T_{kk\PRIME}
      \,
      \ORD{
      \Cadd{k\NOETA}
      \Cdel{k\PRIME\NOETA}
      }
      \,,
    \\
    T_{kk\PRIME}
    &=
      \VF k\,\delta_{kk\PRIME}
    +
    w_{k\PRIME-k}
    +
    (k+k\PRIME\,
    )\frac{g_{k\PRIME-k}}{4\pi}
    \,.\nonumber
  \end{align}%
  We conclude that the four-fermion interaction terms
  in~\eqref{eq:chiralH} cancel, as they do in the Kronig
  identity~\eqref{eq:H01}. In terms of field operators we obtain
  \begin{subequations}\label{eq:chiralHfermionicrealspace}%
    \begin{align}%
      H_{\text{chiral}}
      &=
        \int
        \!\DX\,
        \ORD{
        \frac{
        \PSIadd{\NOETA}{x}
        h(x)
        \PSIdel{\NOETA}{x}
        }{2\pi}
        }
        \,,
      \\
      h(x)
      &=
        \tilde{g}(x)(-\IMI s\partial_x)-\frac12\IMI s\tilde{g}\PRIME(x)+w(x)
        \,,\label{eq:linearkinetic}
    \end{align}%
  \end{subequations}
  where $\tilde{g}(x)$ $=$ $2\pi \VF$ $+$ $g(x)$ as above, and
  $s$ $=$ $-1/(2\pi)$.

  Next we use the spectrum of the first-quantized Hamiltonian
  in~\eqref{eq:linearkinetic}, $h$ $=$
  $s[\tilde{g}(X)P+P\tilde{g}(X)]/2+w(X)$ with $[X,P]$ $=$ $\IMI$.
  The eigenvalue equation $h(x)\xi_k(x)$ $=$ $E_k\xi_k(x)$ is
  separable because $h$ is linear in $P$. For a constant real
  scale~$s$ and real functions $\tilde{g}(x)$, $w(x)$ on an interval
  $[x_1,x_2]$ with $\tilde{g}(x)$ $>$ $0$ and $\tilde{g}(x_1)$ $=$
  $\tilde{g}(x_2)$, and demanding $\xi_k(x_2)$ $=$
  $\xi_k(x_1)\EXP^{\pi\IMI\delta_\text{b}}$, we find $E_k$ $=$ $(S_1$
  $-$ $sLk)/S_0$, $\xi_k(x)$ $=$
  $[\tilde{g}(x)S_0]^{-\frac12}\exp(\IMI[s_0(x,0)E_k$ $-$
  $s_1(x,0)]/s)$, where the momentum $k$ takes on the same discrete
  values $k_n$ as before. Here $S_j$ $=$ $s_j(x_2,x_1)$ with
  $s_j(x,x')$ $=$
  $\int_{x'}^x\DY\,(\delta_{j0}+\delta_{j1}w(y))/\tilde{g}(y)$. These
  eigenstates 
  correspond to plane waves subject to a local scale transformation
  induced by the interaction potential, reminiscient of eikonal wave
  equations or semiclassical Schr\"odinger equations.  We note the
  eigenstate expectation values $\expval{w(X)}$ $=$ $S_1/S_0$ $=$
  $E_k-s\expval{P}$.
  
  Setting
  $x_1$ $=$ $-x_2$ $=L/2$ and $\delta_\text{b}$ $=$ $1$ and requiring
  $g_{q=0}$ $>$ $-2\pi \VF$, we thus
  diagonalize~\eqref{eq:chiralH}, \eqref{eq:chiralHfermionic},
  \eqref{eq:chiralHfermionicrealspace} in terms of new canonical
  fermions, $\{\XIdel{k\NOETA},\XIadd{k\PRIME\NOETA}\}$ $=$
  $\delta_{kk\PRIME}$,
  as
  \begin{align}\label{eq:chiralH_diagonal}%
    H_{\text{chiral}}
    &=
      \sum_kE_k\,\ORD{\XIadd{k\NOETA}\XIdel{k\NOETA}}
    \equiv
    H_{[\bm{\tilde{g}},\bm{w};\bm{\XIdel{\NOETA}}]}^{\text{diagonal}}
    \,,
    \\
    E_k
    &=
      \tilde{v}(k-\tilde{k})
      \,,
      \,~~
      \hatN
      =
      \sum_k
      \ORD{
      \XIadd{k\NOETA}
      \XIdel{k\NOETA}
      }
      =
      \sum_k
      \ORD{
      \Cadd{k\NOETA}
      \Cdel{k\NOETA}
      }
      \,,\nonumber
    \\
    \XIdel{k\NOETA}
    &=
      \int\!\frac{\DX}{\sqrt{2\pi}}\,\xi_k(x)\,
      \PSIdel{\NOETA}{x}
      \,,~~
      \tilde{k}
      =
      -\int\!\frac{\DX}{L}\,\frac{2\pi w(x)}{\tilde{g}(x)}
      \,,\nonumber
    \\
    \xi_k(x)
    &=
      \frac{
      \sqrt{2\pi\tilde{v}}\;
      \EXP^{-\IMI [\tilde{r}_0(x)k-\tilde{r}_1(x)]}
      }{
      \sqrt{L\,\tilde{g}(x)}
      }
      \,,~
      \tilde{r}_0(x)
      =
      \int_0^x\!\DY\,\frac{2\pi\tilde{v}}{\tilde{g}(y)}
      \,,
      \nonumber\\
    \tilde{r}_1(x)
    &=
      \tilde{k}\,\tilde{r}_0(x)
      +
      \int_0^x\!\DY\,\frac{2\pi w(y)}{\tilde{g}(y)}
      ,~
      \tilde{v}
      =
      \bigg[\int\!\frac{\DX}{L}\,\frac{2\pi}{\tilde{g}(x)}\bigg]^{-1}
      \!.\nonumber
  \end{align}
  Note that the renormalized dressed Fermi velocity $\tilde{v}$ is
  given by the spatial harmonic average of the renormalized `local'
  Fermi velocity $\VF$ $+$ $g(x)/(2\pi)$ $=$ $\tilde{g}(x)/(2\pi)$.


  \subsection{Green function for the chiral model}

  From the above solution it is straightforward to obtain the time-ordered
  Green function for the Heisenberg operators of the chiral field,
  \begin{align}
    G(x,x\PRIME;t)
    &=
      \theta(t)\Ggtr(x,x\PRIME;t)
      -
      \theta(-t)\Gles(x,x\PRIME;t)
      \,,
    \\
    \Ggtrles(x,x\PRIME;t)
    &=
      \begin{cases}
        -\IMI\expval{\PSIdel{\NOETA}{x,t}\,\PSIadd{\NOETA}{x\PRIME,0}}
        \,,\\
        -\IMI\expval{\PSIadd{\NOETA}{x\PRIME,0}\,\PSIdel{\NOETA}{x,t}}
        \,,
      \end{cases}
    \label{eq:green_chiral}
  \end{align}
  with $\theta(\pm t)$ $=$ $(1\pm\text{sgn}(t))/2$. At zero
  temperature in a state with fixed particle number $N$ we find
  \begin{align}
    \label{eq:Gchiral_zerotemp}
    \IMI G(x,x\PRIME;t)
    &=
    \frac{
    \tilde{v}
    }{
    \sqrt{
    \tilde{g}(x)\tilde{g}(x\PRIME)
    }
    }
    \frac{
    \EXP^{ 
    \IMI S(x,x',t)
    }
    }{
    \frac{L}{\pi}\sh\frac{\pi}{L}(\IMI R(x,x',t)+a\,\text{sgn}\,t)
    }
    \,,\nonumber
    \\
    R(x,x',t)
    &=
      \tilde{r}_0(x)-\tilde{r}_0(x')-\tilde{v}t
      \,,
    \\
    S(x,x',t)
    &=
      \tilde{r}_1(x)
      -%
      \tilde{r}_1(x')
      +\tilde{v}\tilde{k}t
      -\frac{2\pi N}{L}    R(x,x',t)
      \,,\nonumber
  \end{align}
  where $a$ $\to$ $0^+$ stems from a convergence factor that was
  included in the momentum sum. For constant $g(x)$ and $w(x)$ we
  recover the translationally invariant case, $G(x,x\PRIME;t)$
  $\propto$ $1/(x-x'+\tilde{v}t+a\,\text{sgn}\, t)$, with renormalized
  Fermi velocity.  Position-dependent couplings, on the other hand,
  may lead to a substantial redistribution of spectral weight.  The
  critical behavior however remains unaffected, in the sense that the
  exponent of the denominator involving $R(x,x',t)$ remains unity for
  the chiral model.


  \section{Luttinger droplets}\label{sec:droplet}

  \subsection{Droplet model with with right and left movers}

  We now study a generalization of the two-flavor Tomonaga-Luttinger
  model to position-dependent interactions and scattering potentials.
  Such a `Luttinger droplet' involves right- and left-moving fermions,
  $\PSIdel{\text{R}}{x}$ $=$ $\PSIdel{1}{-x}$ and
  $\PSIdel{\text{L}}{x}$ $=$ $\PSIdel{2}{x}$ (see introduction) with
  linear dispersion in opposite directions, subject to the
  one-particle potential $W(x)$, as well as intrabranch and
  interbranch density interactions $V(x)$ and $U(x)$, respectively, as
  given in~\eqref{eq:dropletHfermionicLR-intro}.  In terms of fermions
  with flavor $\eta$ $=$ $1,2$ we have
  \begin{align}
    &H
      =
      \AST
      \sum_{\eta}
      \Bigg[
      \VF
      \sum_{k}
      k 
      \,
      \Cadd{k\eta}
      \Cdel{k\eta}
      +
      \int
      \!
      \frac{\DX}{L}\,
      W(x)
      n_{\eta}(x) 
      \nonumber\\&~~
    +
    \frac12V(x)n_{\eta}(x)^2      
    \Bigg]
    +
    \int
    \!\frac{\DX}{L}\,
    U(x)
    n_{1}(-x)n_{2}(x)
    \AST
    \,,
  \end{align}%
  i.e., compared to~\eqref{eq:chiralH} the couplings $g(x)$ and $w(x)$ were
  relabeled as $V(x)$ and $W(x)$, indices $\eta$ were put on operators, and the
  interaction term with $U(x)$ was included.

  \subsection{Diagonalization of the Luttinger droplet model}

  \subsubsection{Bosonic form of the Hamiltonian}
  
  Rewritten with bosonic operators
  this becomes
  \begin{align}
    \label{eq:dropletHbosonic}
    H
    &=
      H_\text{TL}+H'+H''
      \,,
    \\
    H_\text{TL}
    &=
      \sum_\eta
      \bigg[
      \frac{2\pi \VF+V_0}{L}
      \bigg(
      \frac{\hatN_{\eta}^2}{2}
      +
      \sum_{q>0}
      K_{-q\eta}^{(0)}
      K_{q\eta}^{(0)}
      \bigg)
      \bigg]
      \nonumber\\&\,\hphantom{=}\,
    +
    \frac{U_0}{L}
    \bigg[
    \hatN_{1}\hatN_{2}
    +
    \sum_{q>0}
    \bigg(
    K_{-q1}^{(0)}
    K_{-q2}^{(0)}
    +
    K_{q1}^{(0)}
    K_{q2}^{(0)}
    \bigg)
    \bigg]
    ,\nonumber
    \\
    H'
    &=
      \sum_{\eta}
      \bigg[
      W_0
      \hatN_{\eta}
      +
      \sum_{q\neq0}
      \bigg(
      W_q
      +  
      \frac{V_q}{L}
      \hatN_{\eta}
      +
      \frac{U_q}{2L}
      \hatN_{\bar{\eta}}
      \bigg)
      K_{q\eta}^{(0)}
      \bigg]
      ,\nonumber
    \\
    H''
    &=
      \sum_{\eta}
      \sum_{q\neq0}
      \sum_{p(\neq0,q)}\!\!
      K_{p\eta}^{(0)}
      \bigg[
      \frac{V_q}{2L}
      K_{q-p\eta}^{(0)}
      +
      \frac{U_q}{2L}
      K_{p-q\bar{\eta}}^{(0)}
      \bigg]
      \,.\nonumber
  \end{align}
  $H$ contains a standard (i.e., translationally invariant)
  Tomonaga-Luttinger model $H_\text{TL}$ involving only the
  zero-momentum (space-averaged) couplings, which by itself can be
  diagonalized by a Bogoljubov transformation. For position-dependent
  couplings, on the other hand, also $H'$ (linear in bosons) and $H''$
  (quadratic in bosons with momentum transfer) are present.
  

  \subsubsection{Specialization to common spatial dependence}

  For simplicity we set
  from now on
  \begin{align}\label{eq:dropletspatialdependence}%
    \binom{V(x)}{U(x)}
    &=
      \binom{V_0}{U_0}
      +
      \binom{V}{U}
      \sum_{q\neq0}f_q\cos(qx)
      \,,
  \end{align}%
  with constant prefactors $V$ and $U$ and  $f_q$ $=$
  $V_q/V$ $=$ $U_q/U$ $=$ $f_{-q}$ for $q$ $\neq$ $0$.
  We can then simplify the momentum-offdiagonal term $H''$ by a Bogoljubov
  transformation to $K_{q\sigma}^{(0)}$ (for $q$ $\neq$ $0$, $\sigma$
  $=$ $-\bar{\sigma}$ $=$ $\pm$, letting $\eta_{\sigma}$ $=$
  $(3$\,$-$\,$\sigma)/2$, $\sigma_\eta$ $=$ $3$\,$-$\,$2\eta$ for
  $\eta$ $=$ $3$\,$-$\,$\bar{\eta}$ $=$ $1,2$),
  \begin{subequations}\label{eq:main:bogoljubov}%
    \begin{align}%
      K_{q\sigma}^{(0)}
      &=
        \COSHTHETA
        \,
        K_{q\eta_{\sigma}}^{(0)}
        +
        \SINHTHETA
        \,
        K_{-q\bar{\eta}_{\sigma}}^{(0)}
        \,,
      \\
      K_{q\eta}^{(0)}
      &=
        \COSHTHETA
        \,
        K_{q\sigma_\eta}^{(0)}
        -
        \SINHTHETA
        \,
        K_{-q\bar{\sigma}_\eta}^{(0)}
        \,,
    \end{align}%
  \end{subequations}%
  $u$ $=$ $\ch\!\theta$,
  $v$ $=$ $\sh\!\theta$,
  which preserves the bosonic algebra,
  $[K_{-q\sigma}^{(0)},K_{q\PRIME\sigma\PRIME}^{(0)}]$ $=$
  $-n_q\delta_{qq\PRIME}\delta_{\sigma\sigma\PRIME}$. The choice
  $U/V$ $=$ $\tnh 2\theta$, assuming $|U|$ $<$
  $V$, yields
  \begin{align}
    \label{eq:dropletHbosonicsigma}
    H
    &=
      \sum_{\sigma=\pm}\big(H_{\sigma}^{(0)}+H_{\sigma}^{(1)}\big)+H^{(2)}+\hatH_\text{N}+E_0
      \,,
    \\
    \!\!\!\!
    H_{\sigma}^{(0)}
    &=
      \frac{\bar{V}}{L}
      \sum_{q>0}
      K_{-q\sigma}^{(0)}
      K_{q\sigma}^{(0)}
      +
      \sum_{q\neq0}
      \frac{\bar{U}\!f_q}{2L}
      \sum_{p(\neq0,q)}\!\!
      K_{p\sigma}^{(0)}
      K_{q-p\sigma}^{(0)}
      \,,\nonumber
    \\
    \!\!\!\!
    H_{\sigma}^{(1)}
    &=
      \frac{1}{L}
      \sum_{q\neq0}
      \hat{\bar{w}}_{q\sigma}
      K_{q\sigma}^{(0)}
      \,,~~~~
      H^{(2)}
      =
      \frac{\bar{V}'}{L}
      \sum_{q\neq0}
      K_{q+}^{(0)}
      K_{q-}^{(0)}
      \,,\nonumber
    \\
    \hatH_{\text{N}}
    &=
      \frac{2\pi \VF+V_0}{2L}\sum_\eta\hatN_\eta^2+\frac{U_0}{L}\hatN_1\hatN_2
      +W_0\sum_\eta\hatN_\eta
  \end{align}
  where $E_0$ is a constant energy shift, omitted from now on, which
  diverges due to the contact interactions in $H$.  Here and below we
  use the following abbreviations and relations,
  \begin{align}\label{eq:ham_par_bos}
    \bar{V}
    &=
      \frac{(2\pi \VF+V_0)V-U_0U}{\bar{U}}
      \,,~~
      \bar{U}
      =
      \bar{\gamma}V
      ,~
    \\
    \bar{V}'
    &=
      \frac{U_0V-(2\pi \VF+V_0)U}{\bar{U}}
      \,,\nonumber
    \\
    \hat{\bar{w}}_{q\sigma}
    &=
      LW_q\EXP^{-\theta}
      +
      \bar{\gamma}\,V_q
      \big[
      \COSHTHETA^3\hatN_1\delta_{\sigma+}
      -
      \SINHTHETA^3\hatN_2\delta_{\sigma-}
      \big]
      ,~(q\neq0)\!\!
      \nonumber
    \\
    \gamma
    &=
      \frac{U}{V}=\tnh\!2\theta
      \,,~
      \bar{\gamma}
      =
      \sqrt{1-\gamma^2}
      =
      \text{sech}{2\theta}
      \,,\nonumber
    \\
    \gamma_3
    &=
      u^3-v^3
      =(1+\tfrac{1}{2}\gamma)(1-\gamma)^{-\frac14}(1+\gamma)^{-\frac34}
      \,,\nonumber
    \\
    \EXP^{-\theta}
    &=
      (V-U)^{\frac14}(V+U)^{-\frac14}
      =
      (1-\gamma)^{\frac14}(1+\gamma)^{-\frac14}
      \,,\nonumber
    \\
    2v^2
    &=
      2\sinh^2\!\theta=(1-\gamma^2)^{-\frac12}-1
      \,.\nonumber
  \end{align}
  The Hamiltonian $H$ has thus become diagonal in the new flavors
  $\sigma$ except for the term $H^{(2)}$ in~\eqref{eq:dropletHbosonicsigma}.
  
  \subsubsection{Specialization to interrelated interaction strengths}

  For simplicity we now assume that $\bar{V}'$ $=$ $0$, i.e., that the
  bare Fermi velocity $\VF$ 
  and the strengths of the position-averaged ($V_0$ and $U_0$) and
  position-dependent interactions ($V$ and $U$) combine so that
  $H^{(2)}$ is absent. This corresponds to the special case
  \begin{align}
    \gamma
    &=
      \frac{U}{V}=\frac{U_0}{2\pi \VF+V_0}
      \,,\label{eq:gamma-specialcase}
  \end{align}
  which together with~\eqref{eq:dropletspatialdependence} is
  equivalent to~\eqref{eq:dropletHfermionicLR-intro-parameters}.  From
  now on we will thus consider $\VF$, $V_q$, $\gamma$ to be chosen
  freely (with $V_0$ $>$ $-2\pi\VF$), with the other parameters in $H$
  then being given by
  \begin{subequations}%
    \begin{align}%
      U_q&=\gamma
           \,
           (2\pi \VF\delta_{q0}+V_q)
           \,,
      \\
      \bar{V}
         &=
           \bar{\gamma}
           \,
           (2\pi \VF+V_0)
           \,,~
           \bar{U}
           =
           \bar{\gamma}
           \,
           V
           \,,~
           \bar{V}'
           =
           0
           \,,~
    \end{align}%
  \end{subequations}%
  i.e., $\bar{U} f_q$ $=$ $\bar{\gamma} V_q$ for $q$ $\neq$ $0$.
  Then for $\sigma$ $=$ $\pm1$ each decoupled Hamiltonian has precisely
  the form of the bosonic Hamiltonian~\eqref{eq:chiralHbosonic}
  encountered in the chiral model,
  \begin{align}\label{eq:dropletHbosonic-sigmas}%
    H
    &=
      \hatH_\text{N}
      +
      \sum_{\sigma=\pm}
      {H}_{\sigma}
      \,,\\
    H_{\sigma}
    &
      =
      H_{\sigma}^{(0)}+H_{\sigma}^{(1)}
      =
      H_{[\bar{\bm{g}},\hat{\bar{\bm{w}}}_\sigma;\bm{K}_\sigma]}^{\text{bosonic}}
      \,,\nonumber
  \end{align}
  with effective interaction $\bar{g}_{q}$ $=$
  $\bar{V}\delta_{q0} + (1-\delta_{q0})\bar{U}f_q$,
  i.e.,
  \begin{subequations}%
    \begin{align}
      \bar{g}_{q}
      &=
        \bar{\gamma}
        \,
        [2\pi \VF\delta_{q0}+V_q]      
        \,,
        ~~
        \bar{g}(x)
        =
        \bar{\gamma}
        \,
        [2\pi \VF+V(x)]
        \,,
      \\
      \bar{v}
      &=
        \bigg[\int\!\frac{\DX}{L}\,\frac{2\pi}{\bar{g}(x)}\bigg]^{-1}
        \,,\label{eq:vbar+Wbar-def}
        ~~
        \bar{W}
        =
        \int\!\frac{\DX\,2\pi\bar{v}}{L}\,\frac{W(x)}{\bar{g}(x)}
        \,,
    \end{align}%
  \end{subequations}%
  where we also introduced the renormalized Fermi velocity $\bar{v}$
  and averaged one-particle potential $\bar{W}$ which will emerge below.
  
  \subsubsection{Refermionization as separately diagonalizable chiral models}

  We thus refermionize each $H_\sigma$, first in terms of new
  fermions $\PSIdel{\sigma}{x}$, with bosonic fields $\phi_\sigma(x)$ $=$
  $\PHIadd{\sigma}{x}$ $+$ $\PHIdel{\sigma}{x}$ built from the
  $K_{q\sigma}^{(0)}$ analogously to~\eqref{eq:intro:bosonizationidentity},
  \begin{align}
    \PSIdel{\sigma}{x}
    &=
      \sqrt{\frac{2\pi}{L}}\;
      \sum_{k} \EXP^{-\IMI kx} \Cdel{k\sigma}
      \nonumber\\
    &=
      \frac{\Fdel{\sigma}}{\sqrt{a}}
      \,
      \EXP^{-\IMI\frac{2\pi}{L}(\hatN_{\sigma}-\frac12)x}
      \,
      \EXP^{-\IMI\phi_\sigma(x)}
      \,.
      \label{eq:refermionization:sigma}
  \end{align}
  Below we will fix the connection between the fermionic number
  operators $\hatN_\sigma$ and their associated Klein factors
  $\Fdel{\sigma}$ to the original fermions $\Cdel{k\eta}$, which is
  not determined by the purely bosonic Bogoljubov
  transformation~\eqref{eq:main:bogoljubov}.

  Next each chiral-type Hamiltonian $H_\sigma$ is diagonalized with
  fermions $\Xi_{k\sigma}$ according to~\eqref{eq:chiralH_diagonal},
  \begin{align}\label{eq:X-refermionize}
    H_\sigma
    =
    H_{[\bar{\bm{g}},\hat{\bar{\bm{w}}}_\sigma;\bm{K}_\sigma]}^{\text{bosonic}}
    &=
      H_{[\bar{\bm{g}},\hat{\bar{\bm{w}}}_\sigma;\bm{\Xi}_{\sigma}]}^{\text{diagonal}}
      -
      \frac{\bar{g}_0}{2L}\hatN_{\sigma}^2
      \,,
    \\
    H_{[\bar{\bm{g}},\hat{\bar{\bm{w}}}_\sigma;\bm{\Xi}_{\sigma}]}^{\text{diagonal}}
    &=
      \bar{v}\sum_k(k-\hat{\bar{k}}_{\sigma})\ORD{\XIadd{k\sigma}\XIdel{k\sigma}}
      \,.
      \nonumber
  \end{align}
  with the two types of fermions $\psi_\sigma$ and $\Xi_\sigma$
  related by
  \begin{subequations}%
    \begin{align}%
      \XIdel{k\sigma}
      &=
        \int\!\frac{\DX}{\sqrt{2\pi}}\,\xi_{k\sigma}(x)\,
        \PSIdel{\sigma}{x}
        \,,
        \label{eq:Xi-k-sigma-definition}
      \\
      \hatN_\sigma
      &=
        \sum_k\ORD{\XIadd{k\sigma}\XIdel{k\sigma}}
        = \sum_k \ORD{\Cadd{k\sigma} \Cdel{k\sigma}}
        \label{eq:Nsigma-densities}
        \,,
    \end{align}%
  \end{subequations}%
  in terms of the following functions and parameters
  \begin{subequations}\label{eq:xi+r0}%
    \begin{align}%
      \xi_{k\sigma}(x)
      &=
        \frac{
        \sqrt{2\pi\bar{v}}\;
        \EXP^{-\IMI [r_0(x)k-\hat{r}_{1\sigma}
        (x)]}
        }{
        \sqrt{L\,\bar{g}(x)}
        }
        \,,
      \\
      r_0(x)
      &=
        \int_0^x\!\DY\,\frac{2\pi\bar{v}}{\bar{g}(y)}
        =
        -r_0(-x)
        \,,
      \\
      \hat{r}_{1\sigma}(x)
      &=
        \hat{\bar{k}}_{\sigma}\,r_0(x)
        +
        \int_0^x\!\DY\,\frac{2\pi \hat{\bar{w}}_{\sigma}(y)}{\bar{g}(y)}
        \,,
      \\
      \hat{\bar{k}}_{\sigma}
      &=
        -\int\!\frac{\DX}{L}\,\frac{2\pi}{\bar{g}(x)}\sum_{q\neq0}\frac{\hat{\bar{w}}_{q\sigma}}{L}\EXP^{-\IMI qx}
        \,.
    \end{align}%
  \end{subequations}%

  \subsubsection{Rebosonization into canonical form with quadratic number operator terms}

  Due to the linear dispersion we can rebosonize the $\Xi_{k\sigma}$
  in terms of new canonical bosons $\BIGBdel{q\sigma\!}$, which will
  also be needed for the calculation of Green
  functions below.  The corresponding (re-)bosonization identity
  reads
  \begin{align}
    \!\!\!
    \Xi_{\sigma}(x)
    &=
      \frac{\calFdel{\sigma}}{\sqrt{a}}
      \EXP^{-\frac{2\pi\IMI}{L}(\hatN_\sigma-\frac12)x
                       +\sum\limits_{q>0}\!\!
                       \frac{\IMI\EXP^{-a|q|}}{\sqrt{n_q}}
                       \![
                       \LBdel{q\sigma}\EXP^{-\IMI qx}
                       +
                       \LBadd{q\sigma}\EXP^{\IMI qx}
                       ]\!
                       },\!\!\!\label{eq:Xi_rebosonization}
  \end{align}
  where $\calFdel{\sigma}$ is another Klein factor which lowers
  $\hatN_\sigma$ by one. We note that once we fix $\Fdel{\sigma}$,
  then $\calFdel{\sigma}$ is determined
  by~\eqref{eq:refermionization:sigma},
  \eqref{eq:Xi-k-sigma-definition}, \eqref{eq:Xi_rebosonization},
  although its explicit form is not needed in the following. The
  transformation~\eqref{eq:Xi_rebosonization} yields
  \begin{align}%
    \label{eq:droplet-but-numberoperators}
    H
    &=
      H_{\Xi}+H_\text{N}
      -
      \sum_\sigma
      \bigg(
      \frac{\bar{g}_0}{2L}\hatN_\sigma^2
      +
      \bar{v}
      \bar{k}_{\sigma}\hatN_{\sigma}
      \bigg)
      ,\!\!
    \\
    H_{\Xi}
    &=
      \sum_{\sigma;k}\bar{v}k\,\ORD{\XIadd{k\sigma}\XIdel{k\sigma}}
    =
    \sum_{\sigma;q>0}\bar{v}q\BIGBadd{q\sigma}\BIGBdel{q\sigma}
    +
    \frac{\pi}{2L}\sum_\sigma\hatN_\sigma^2
    ,\nonumber
  \end{align}
  We observe that even for position-dependent interactions,
  collective bosonic excitations with linear dispersion emerge.

  To complete the diagonalization of $H$
  in~\eqref{eq:droplet-but-numberoperators}, we must still define the
  new number operators $\hatN_\sigma$ (with integer eigenvalues) and
  Klein factors $\Fdel{\sigma}$ in terms of the original
  $\hatN_{\eta}$ and $\Fdel{\eta}$ (which also appear in
  $H_\text{N}$). We set
  \begin{align}%
    \hatN_\sigma
    &=
      \hatN_1\delta_{\sigma-}
      +
      \hatN_2\delta_{\sigma+}
      \,,
  \end{align}%
  which ensures that the ground state (without bosonic excitations
  $\BIGBadd{q\sigma\!}$) remains in a sector with finite $\hatN_1$
  $=$ $\hatN_2$, because then only the density terms
  $(\hatN_1^2+\hatN_2^2)$ and $\hatN_1\hatN_2$ appear in the
  Hamiltonian.  We note that no other form of $\hatN_\sigma$ that
  is linear in $\hatN_1$ and $\hatN_2$ has this feature The
  corresponding Klein factors are then given by
  \begin{align}%
    \Fdel{\sigma}
    &=
      \Fdel{\mathrm{1}}\delta_{\sigma-}
      +
      \Fdel{\mathrm{2}}\delta_{\sigma+}
      \,.
  \end{align}%
  
  Collecting terms, the diagonalization of the Luttinger droplet
  Hamiltonian~\eqref{eq:dropletHfermionicLR-intro} is then finally
  complete,
  \begin{align}
    \!\!H
    &=
      \sum_{\sigma;q>0}\bar{v}q\BIGBadd{q\sigma}\BIGBdel{q\sigma}
      +
      \frac{\pi}{2L}\big[v_{\calN}\calhatN^2+v_{\calJ}\calhatJ^2]
      +
      \epsilon\calhatN
      \,,\!\!\!\!
      \,,\label{eq:luttingerdroplet-diagonalized}
    \\
    \calhatN
    &=
      \hatN_1+\hatN_2
      \,,~~
      \calhatJ
      =
      \hatN_1-\hatN_2
      \,,\nonumber
  \end{align}
  in which the following parameters appear,
  \begin{align}
    v_{\calN}
    &=
      v_1+v_2
      \,,
    &
      v_{\calJ}
    &=v_1-v_2
      \,,\nonumber
    \\
    v_1
    &=
      \tilde{v}_\text{F}+\Delta v
      \,,
    &
      \tilde{v}_\text{F}
    &=\VF+\tfrac{1}{2\pi}V_0
      \,,\nonumber
    \\
    v_2
    &=
      \gamma\tilde{v}_\text{F}+\gamma_3\Delta v\,,
    &
      \Delta v&=\bar{v}-\bar{\gamma}\tilde{v}_\text{F}
                \,,\nonumber
    \\
    \epsilon
    &=\bar{W}\EXP^{-\theta}
      +W_0(1-\EXP^{-\theta})
      \,,\label{eq:luttingerdroplet-velocities}
      \!\!\!\!\!\!\!\!\!\!\!\!\!\!\!\!\!\!\!\!\!
  \end{align}%
  and $\bar{v}$ and $\bar{W}$ were defined
  in~\eqref{eq:vbar+Wbar-def}.
  Here the total and relative fermionic number operators, $\calhatN$
  and $\calhatJ$, take on integer values and commute with the two
  flavors of bosonic operators.  We note the ground-state value of
  $\calhatN$ may shift due to the one-particle potential $W(x)$
  according to the value $\epsilon$, which also depends on the
  interaction via $\bar{W}$.
   
  We consider~\eqref{eq:luttingerdroplet-diagonalized} to be the
  canonical form of the diagonalized Luttinger droplet Hamiltonian, as
  it is essentially the same as that of the bosonized translationally
  invariant Tomonaga-Luttinger model. Namely, both are characterized
  by the renormalized Fermi velocity $\bar{v}$ for collective bosonic
  particle-hole excitations with linear dispersion, as well as
  $v_{{\calN},{\calJ}}$ for total and relative particle number
  changes.  For the Luttinger droplet, however, spatial dependencies
  enter into the diagonalization and lead to qualitatively different
  behavior for the fermionic degrees of freedom, as discussed below.

  \subsection{Spectrum of the Luttinger droplet model}

  \subsubsection{Recovery of the translationally invariant case}

  For position-independent potentials, the
  translationally invariant case is fully recovered by setting
  $f_{q\neq0}$ $=$ $0$, so that $\bar{v}$ $=$
  $\bar{\gamma}\tilde{v}_\text{F}$ and $\Delta v$ $=$ $0$. We thus
  find that
  \begin{subequations}%
    \begin{align}%
      \!\!\!\!\!\!\!\!\!\!
      W(x)&=W_0\,,~
      V(x)=V_0\,,~
        U(x)=U_0
        \nonumber
      \\[1ex]
      \!\!\!\!\!\!\!\!\!\!
      \Rightarrow~~
      H&=H_\text{TL}+W_0\calhatN
         \,,~
         \gamma= \frac{U_0}{2\pi\VF+V_0}
         \,,
      \\
      v_{{\calN},{\calJ}}
      &=
        \VF+\frac{V_0\pm U_0}{2\pi}
        \nonumber\\&
        =
        \bar{v}
        \bigg[
        \frac{1+\gamma}{1-\gamma}
        \bigg]^{\!\pm\frac12}
        \,,
      \\
      \bar{v}&=
               \sqrt{\Big(\VF+\frac{V_0}{2\pi}\Big)^2-\Big(\frac{U_0}{2\pi}\Big)^2}
               \nonumber\\&
      =
      \bar{\gamma}\Big(\VF+\frac{V_0}{2\pi}\Big)
      =
      \bar{\gamma}\tilde{v}_\text{F}
      \,,
    \end{align}%
    \label{eq:translationallyinvariantcase}%
  \end{subequations}%
  i.e., the parameter $\gamma$ of~\eqref{eq:gamma-specialcase} only
  relates $\VF$, $V_0$, $U_0$ to one another, as the interactions $V$
  and $U$ are absent for the translationally invariant case. As
  before, $\gamma$ characterizes the relative strength of
  (translationally invariant) interbranch interactions.  It is one of
  the characteristic properties of a Luttinger
  liquid\cite{haldane_luttinger_1981} that the relations
  \begin{align}
    \bar{v}
    &=
      \sqrt{v_{\calN}v_{\calJ}^{\vphantom{x}}}
    \,,
    \label{eq:luttingerliquid-vrelation}
    &
    \gamma
    &=
      \frac{v_{\calN}-v_{\calJ}}{v_{\calN}+v_{\calJ}}
      \,,
  \end{align}
  remain valid even if the dispersion in $H_\text{TL}$ is weakly
  nonlinear. This connects the excitation velocities $\bar{v}$,
  $v_{\calN}$, $v_{\calJ}$ as well as the power-law exponents in the
  single-particle Green function, which contain the parameter
  $\gamma$, as discussed below.

  \subsubsection{Excitation velocities for position-dependent interactions}\label{subsubsec:velocities}
  
  By contrast, for the Luttinger
  droplet~\eqref{eq:dropletHfermionicLR-intro} with position-dependent
  interactions, the renormalized Fermi velocity $\bar{v}$ depends on
  $V(x)$ according to~\eqref{eq:vbar+Wbar-def}, so that $\bar{v}$ can
  be varied independently from the average interaction potential
  $V_0$.  Namely if $\bar{v}$ $\neq$ $\bar{\gamma}\tilde{v}_\text{F}$
  in~\eqref{eq:luttingerdroplet-velocities}, i.e., if
  \begin{align}
    \int\!\frac{\DX}{2\pi\VF+V(x)}
    &\neq
      \int\!\frac{\DX}{2\pi\VF+V_0}
      \,,\label{eq:vbar-indep-condition}
  \end{align}
  the three velocities $\bar{v}$, $v_{\calN}$, $v_{\calJ}$ are
  independent of each other (but together determine
  $\gamma$).

  In the following, however, we will adopt a different perspective. We
  regard $\gamma$ as given by the interactions
  as in~\eqref{eq:dropletHfermionicLR-intro-parameters},
  \begin{align}
    \gamma&= \frac{U_0}{2\pi\VF+V_0}= \frac{U(x)}{2\pi\VF+V(x)}
            \,.\label{eq:luttingerdroplet-gamma}
  \end{align}
  Then it follows from~\eqref{eq:luttingerdroplet-velocities} that the
  velocities are related by
  \begin{subequations}%
    \begin{align}%
      \!\!\!
      \bar{v}
      =
      \frac{\gamma-\bar{\gamma}\gamma_3-(1-\bar{\gamma})}{2(\gamma-\gamma_3)}
      &v_{{\calN}}
        +
        \frac{\gamma-\bar{\gamma}\gamma_3+(1-\bar{\gamma})}{2(\gamma-\gamma_3)}
        v_{{\calJ}}
        \,,\!\!
      \\
      \!\!\!
      \tilde{v}_{\text{F}}
      =
      \frac{1-\gamma_3}{2(\gamma-\gamma_3)}
      &v_{{\calN}}
        -
        \frac{1+\gamma_3}{2(\gamma-\gamma_3)}
        v_{{\calJ}}
        \,,
    \end{align}%
    \label{eq:luttingerdroplet-vrelation}%
  \end{subequations}%
  which replaces~\eqref{eq:luttingerliquid-vrelation}.

  Hence we may already conclude that the Luttinger
  droplet~\eqref{eq:dropletHfermionicLR-intro} is strictly speaking
  \emph{not} a Luttinger liquid, in the sense that $\bar{v}$ $\neq$
  $\sqrt{v_{\calN}v_{\calJ}}$ if~\eqref{eq:vbar-indep-condition}
  holds, so that the Luttinger liquid
  relation~\eqref{eq:luttingerliquid-vrelation} is violated and the
  linear relations~\eqref{eq:luttingerdroplet-vrelation} between the
  velocities $\bar{v}$, $v_{{\calN}}$, $v_{{\calJ}}$,
  $\tilde{v}_{\text{F}}$ hold instead.

  Note also that while the canonical form of the
  Hamiltonian~\eqref{eq:luttingerdroplet-diagonalized} and its
  eigenvalues are very similar to the translationally invariant case,
  their relation to the original fermions is more complex since it was
  obtained from a position-dependent canonical transformation.  As a
  result, the position dependence of the interaction appears in the
  Green function, which we calculate next.

  \subsection{Green function for Luttinger droplet model}\label{sub:dropletgreen}
  
  \subsubsection{Rebosonization route to the Green function}

  As in the translationally invariant case, the Green function is
  obtained from the bosonization
  identity~\eqref{eq:intro:bosonizationidentity} and the Bogoljubov
  transformation~\eqref{eq:main:bogoljubov}, but also makes use of
  the refermionization~\eqref{eq:refermionization:sigma} and the
  rebosonization~\eqref{eq:Xi_rebosonization}. Using $\phi_\eta(x)$
  $=$ $\PHIadd{\eta}{x}$ $+$ $\PHIdel{\eta}{x}$ $=$
  $u\phi_{\sigma_\eta}(x)$ $+$ $\phi_{\bar{\sigma}_\eta}(-x)$, we
  have
  \begin{align}
    \PSIdel{\eta}{x}
    &=
      \frac{1}{\sqrt{a}}\,\Fdel{\eta}
      \,
      \EXP^{-\IMI\frac{2\pi}{L}(\hatN_{\eta}-\frac12)x}
      \,
      \EXP^{-\IMI[
      \COSHTHETA\phi_{\sigma_\eta\!}(x)
      +
      \SINHTHETA\phi_{\bar{\sigma}_\eta\!}(-x)
      ]}
      \,.
      \label{eq:bosonizationidentity-eta-sigma}
  \end{align}
  To evaluate correlation functions of this field, we need to
  express it in the diagonalizing fermionic basis~\eqref{eq:Xi-k-sigma-definition}.
  We define the auxiliary functions
  \begin{align}
    \lambda_{q}(x)
    &=
      \IMI \frac{\EXP^{-\IMI q x -a |q|/2}}{n_q}
      \,,
      ~
      \tilde{\lambda}(x-x')
      =
      \sum_{q\neq0}\lambda_{q}(x)\EXP^{\IMI q x'}
      \,,
      \nonumber\\
    \tilde{\lambda}(x)
    &=
      \IMI
      \sum_{q\neq0}
      \frac{\EXP^{-\IMI q x -a |q|/2}}{n_q}
      =2\sum_{n=1}^\infty\frac{1}{n}\sin\frac{2\pi n x}{L}
    \\&
    =
    \pi\,\text{sgn}(x)-\frac{2\pi x}{L}
    \,,
    ~~~~(-L<x<L)        
    \nonumber
  \end{align}
  in terms of which we can express the bosonic fields as
  \begin{align}
    \phi_{\sigma}(x)
    &=
      \sum_{q\neq0}
      \lambda_{q}(x)
      K_{q\sigma}^{(0)}
      =
      \sum_{q \neq 0,k}
      \lambda_{q}(x)
      \Cadd{k-q\sigma}\Cdel{k\sigma}
      \nonumber\\&
    =
    \int\!\frac{\DX'}{2\pi}\,\tilde{\lambda}(x-x')\,\ORD{\PSIadd{\sigma}{x'}\PSIdel{\sigma}{x'}}
    \\&
    =
    \sum_{k,k'}
    \chi_{k-k'}(x)
    \ORD{\XIadd{k\sigma}\XIdel{k'\sigma}}
    \,.
    \nonumber
  \end{align}
  Here further auxiliary functions were introduced,
  \begin{align}
    \chi_{q}(x)
    &=
      2\pi\bar{v}
      \int\!\frac{\DY}{L}\,
      \frac{\tilde{\lambda}(x-y)}{\bar{g}(y)}
      \,
      \EXP^{-\IMI qr_0(y)}
      \nonumber\\&
    =
    \frac{2\pi \IMI}{qL}
    \big(\EXP^{-\IMI qr_0(x)}
    -\bar{R}_q\big)
    \,,
    ~~
    \chi_{0}(x)
    =
    \frac{2\pi}{L}r_0(x)
    \,,
    \label{eq:chiq-def}
    \\
    \bar{R}_q
    &=
      \frac{2}{L}\int_0^{L/2}\!\DX\,\cos(qr_0(x))
      \nonumber\\&
    =
    \frac{2}{L}\int_0^{L/2}\!\DR\,x_0'(r)\cos(qr)
    \,,~\bar{R}_0=1\,,
    \label{eq:Rq-def}
    \\
    x_0(r)
    &=
      r+2\sum_{q>0}\bar{R}_q\frac{\sin qr}{q}
      \,\label{eq:x0-from-Rq}
      \,,
  \end{align}
  where $x_0(r)$ is the unique inverse function of $r_0(x)$,
  which was substituted in the integral in~\eqref{eq:Rq-def} and
  expressed in terms of $\bar{R}_q$ via Fourier transform
  in~\eqref{eq:x0-from-Rq} for later reference.
  The rebosonization relation~\eqref{eq:Xi_rebosonization} then yields
  \begin{align}
    \phi_{\sigma}(x)
    &=
      \sum_{q}
      \chi_{-q}(x)
      \sum_{k'}
      \ORD{\XIadd{k'-q\sigma}\XIdel{k'\sigma}}
      \nonumber\\
    &=
      \chi_{0}(x)\hatN_{\sigma}
      + \IMI A_\sigma(x)
      \,,\\
    A_\sigma(x)
    &=
      \sum_{q>0}\chi_{-q}(x)\sqrt{n_q}\LBdel{q\sigma}
      -\text{h.c.}
      \,,
  \end{align}
  finally expressing the fermionic
  field~\eqref{eq:bosonizationidentity-eta-sigma} in the diagonal
  bosonic basis~\eqref{eq:luttingerdroplet-diagonalized}.  For the
  Green function we also need the time dependence of the Klein
  factors, which originates from $H_\text{N}+H'$ in
  \eqref{eq:dropletHbosonic} and \eqref{eq:dropletHbosonicsigma}.
  This leads to a sum over $K_{q\sigma_\eta}^{(0)}$ which we
  calculate from the inversion $K_{q\sigma}^{(0)}$ $=$
  $\int\!\frac{\DX}{2\pi}\, \EXP^{-\IMI q x}
  \partial_x\phi_{\sigma}(x)$ ($q$ $\neq$ $0$), namely
  $\sum_{q\neq0} V_q K_{q\sigma}^{(0)}$ $=$
  $\bar{\kappa}_{0}\hatN_{\sigma}/L$ $+$ $\IMI \bar{A}_\sigma$,
  where
  \begin{align}
    \bar{A}_\sigma
    &=
      \sum_{q>0}
      \bar{\kappa}_{-q}\sqrt{n_q}\LBdel{q\sigma}
      -
      \text{h.c.}
      \,,
    \\
    \bar{\kappa}_{q}
    &=
      \int\!\frac{\DX}{2\pi L}\,
      (V(x)-V_0)
      \chi_{q}'(x)
      =
      -\frac{2\pi\tilde{v}_\text{F}}{L}\bar{R}_q
      \,.
  \end{align}
  Using the hyperbolic relation $\EXP^{\mp\theta}(1\pm\gamma/2)$ $=$
  $\bar{\gamma}(u^3\mp v^3)$ and eliminating $U_0$
  with~\eqref{eq:gamma-specialcase}, the time-dependent Klein factor
  then becomes
  \begin{align}
    \Fdel{\eta}(t)
    &=
      \EXP^{\IMI (H_\text{N}+H')t}
      \Fdel{\eta}\,
      \EXP^{-\IMI (H_\text{N}+H')t}
      \nonumber\\&
    =
    \Fdel{\eta}
    \EXP^{
    -\IMI t
    [
    2\pi\tilde{v}_\text{F}(\hatN_\eta+\gamma\hatN_{\bar{\eta}}-\frac12)/L+W_0
    +
    \bar{\gamma}\bar{\kappa}_0(
    u^3\hatN_{\bar{\eta}}
    -
    v^3\hatN_{\eta})]
    }\nonumber\\&~~~~~~~~\times\EXP^{
    \IMI\bar{\gamma}(
    u^3
    \bar{A}_{\sigma_\eta}
    -
    v^3\bar{A}_{\bar{\sigma}_\eta}
    )
    }
    \,.
  \end{align}
  We evaluate the Green function in the ground state with $\hatN_\eta$
  $=$ ${\calN}/2$ $=$ $\hatN_\sigma$ and
  $\LBadd{q\sigma}\LBdel{q\sigma}$ $=$ $0$ for all $q$ $>$ $0$, where
  $\calN$ is the integer closest to $-\epsilon/(2v_{\cal N})$,
  \begin{align}
    G_\eta(x,x\PRIME;t)
    &=
      \theta(t)\Ggtr_{\eta\eta}(x,x\PRIME;t)
      -
      \theta(-t)\Gles_{\eta\eta}(x,x\PRIME;t)
      \,,
  \end{align}
  with $\theta(\pm t)$ $=$ $(1\pm\text{sgn}(t))/2$. 
  The greater and lesser Green functions,
  \begin{align}
    \Ggtrles_{\eta\eta'}(x,x';t)
    &=
      \begin{cases}
        -\IMI\expval{\PSIdel{\eta}{x,t}\,\PSIadd{\eta'}{x',0}}
        \,,\\
        -\IMI\expval{\PSIadd{\eta'}{x',0}\,\PSIdel{\eta}{x,t}}
        \,,
      \end{cases}
    \nonumber\\&
    =
    \delta_{\eta\eta'}
    \Ggtrles_{\eta\eta}(x,x';t)
    \,,
  \end{align}
  are then flavor-diagonal.  They are evaluated by first clearing the
  Klein factors, inserting the Bogoljubov-transformed bosonic fields,
  separate them according to the index $\sigma$, and then express them
  with $\hatN_\sigma$, $\LBdel{q\sigma}$, $\LBadd{q\sigma}$. This
  leads to
  \begin{align}
    &\IMI a\Ggtrles_{\eta}(x,x\PRIME;t)
      =
      M^{\gtrless}_{x,x',t}
      \,
      M_{\sigma_\eta}(t\bar{\gamma}u^3,\pm u,x_{\gtrless},t_{\gtrless},x_{\lessgtr},t_{\lessgtr})
      \nonumber\\&~~~~~~\times
    M_{\bar{\sigma}_\eta}(-t\bar{\gamma}v^3,\pm v,-x_{\gtrless},t_{\gtrless},-x_{\lessgtr},t_{\lessgtr})
    \,,    \label{eq:Gdroplet-result}
  \end{align}
  with a phase factor and $\sigma$-diagonal exponential bosonic expectation values
  \begin{align}%
    &M^{\gtrless}_{x,x',t}
      =
      \EXP^{-\frac{\IMI\pi}{L}[(\calN\pm1)(x-x')+v_{\gtrless}t]-\IMI\EXP^{-\theta}[\chi_0(x)-\chi_0(x')]\frac{\calN}{2}}
      \!,
      \nonumber\\
    &M_{\sigma}(\tau{},\nu,x,t,x',t')
      =
      \langle
      \EXP^{\tau{}\bar{A}_{\sigma}}
      \EXP^{\nu A_{\sigma}(x,t)}
      \EXP^{-\nu A_{\sigma}(x',t')} 
      \rangle_{\sigma}
      ,
      \label{eq:Msigma-def}
  \end{align}%
  with $x_>$ $=$ $x$, $x_<$ $=$ $x'$, $t_>$ $=$ $t$, $x_<$ $=$ $0$,
  and a velocity parameter given by $v_{\gtrless}$ $=$
  $(\tilde{v}_\text{F}$ $-$ $\bar{\gamma}\kappa_0v^3/(2\pi))(\calN$
  $+$ $1$ $\pm$ $1)$ $+$ $(\tilde{v}_\text{F}\gamma$ $+$
  $\bar{\gamma}\kappa_0u^3/(2\pi))\calN$ $+$ $LW_0/\pi$ $-$
  $\tilde{v}_\text{F}$.  To evaluate the remaining expectation value,
  we use the identity\cite{von_delft_bosonization_1998}
  \begin{align}
    \expval{\EXP^{A_1}\EXP^{A_2}\EXP^{A_3}} = \EXP^{\expval{A_1 A_2+A_2 A_3 + A_1 A_3 + \tfrac{1}{2}(A_1^2+A_2^2+A_3^2)}}\,,
  \end{align}  
  valid for linear bosonic  operators $A_1$, $A_2$, $A_3$ and  eigenstates of the bosonic particle numbers.
  We obtain
  \begin{align}\label{eq:Msigma-result}
    &M
      (\tau{},\nu,x,t,x',t')
    =
    \EXP^{
    -\tfrac{1}{2}\tau^2\bar{S}_0^{[a]}
    -
    \tau\nu
    \big(\bar{S}_1^{[\bar{v}t,a]}(x)-\bar{S}_1^{[\bar{v}t'\!,a]}(x')\big)
    }\nonumber
    \\&~~\times\EXP^{
    \frac{1}{2}\nu^2\big(
    2\bar{S}_2^{[\bar{v}(t'-t),a]}(x,x')
    -\bar{S}_2^{[0,a]}(x,x)
    -\bar{S}_2^{[0,a]}(x',x')
    \big)
    }
    \,,
  \end{align}  
  where the index $\sigma$ was omitted because $M_\sigma$ is independent of
  it, and we used the abbreviations
  \begin{subequations}\label{eq:Sfunc-def}%
    \begin{align}%
      \bar{S}_0
      &=
        \sum_{q>0}
        n_q|\bar{\kappa}_{q}|^2\EXP^{iqs}\EXP^{-aq}
        \,,
      \\
      \bar{S}_1^{[s,a]}(y)
      &=
        \sum_{q>0}
        n_q\bar{\kappa}_{-q}\chi_{q}(y)\EXP^{iqs}\EXP^{-aq}
        \,,
      \\
      \bar{S}_2^{[s,a]}(x,y)
      &=
        \sum_{q>0}
        n_q\chi_{-q}(x)\chi_{q}(y)\EXP^{iqs}\EXP^{-aq}
        \,.
    \end{align}%
  \end{subequations}
  Using the explicit wave functions and the
  definition~\eqref{eq:Rq-def}, they evaluate to
  \begin{subequations}\label{eq:Sfunc-result}%
    \begin{align}%
      \bar{S}_0
      &=
        \Big(\frac{2\pi\tilde{v}_\text{F}}{L}\Big)^2
        \bar{R}_{1,2}^{[0,a]}
        \,,
      \\
      \bar{S}_1^{[s,a]}(y)
      &=
        \frac{2\pi\tilde{v}_\text{F}}{L}
        \IMI
        \Big(
        \bar{R}_{0,2}^{[s,a]}
        -
        \bar{R}_{0,1}^{[s-r_0(y),a]}
        \Big)
        \,,
      \\
      \bar{S}_2^{[s,a]}(x,y)
      &=
        \bar{R}_{-1,0}^{[s+r_0(x)-r_0(y),a]}
        +
        \bar{R}_{-1,2}^{[s,a]}
        \nonumber\\&~~~~~~
      -
      \bar{R}_{-1,1}^{[s+r_0(x),a]}
      -
      \bar{R}_{-1,1}^{[s-r_0(y),a]}
      \,.
    \end{align}
  \end{subequations}
  Here we introduced the functions
  \begin{align}
    \bar{R}_{m,n}^{[s,a]}
    &=
      \sum_{q>0}
      n_q^m\bar{R}_q^n\EXP^{\IMI qs}\EXP^{-aq}
      \,,\label{eq:Rfunc-def}%
  \end{align}
  which for $n$ $\neq$ $0$ depend on the position dependence of $V(x)$
  through $\bar{R}_q$ of~\eqref{eq:Rq-def}.
  Putting~\eqref{eq:Msigma-def}, \eqref{eq:Msigma-result},
  \eqref{eq:Sfunc-result} into~\eqref{eq:Gdroplet-result}, the
  calculation of the Green function is complete, and can be summarized as 
  \begin{subequations}%
    \begin{align}%
      &\Ggtr_{\eta}(x,x\PRIME;t)=M^{>}_{x,x',t}
      \\&~~\times
      M(t\bar{\gamma}u^3,+u,x,t,x',0)
      M(-t\bar{\gamma}v^3,+v,-x,t,-x',0)
      \,,\nonumber
      \\
      &\Gles_{\eta}(x,x\PRIME;t)=M^{<}_{x,x',t}
      \\&~~\times
      M(t\bar{\gamma}u^3,-u,x',0,x,t)
      M(-t\bar{\gamma}v^3,-v,-x',0,-x,t)
      \,,\nonumber
    \end{align}%
  \end{subequations}%
  with the factors given by~\eqref{eq:Msigma-def} and~\eqref{eq:Msigma-result}.
  We now discuss this result for
  different settings, referring for simplicity only to $\Ggtr_{\eta}(x,x\PRIME;t)$.

  \subsubsection{Recovery of the translationally invariant case}

  In the translationally invariant
  case~\eqref{eq:translationallyinvariantcase} we have $r_0(x)$ $=$
  $x$, due to the constant function $r_0'(x)$ $=$
  $\bar{v}/(\bar{\gamma}\tilde{v}_\text{F})$ $=$ $1$, cf.~\eqref{eq:xi+r0}. Also
  $\bar{R}_q$ $=$ $\delta_{q0}$, so that all sums over $\bar{R}_q$ (with
  $q$ $>$ $0$) vanish. In   $\bar{S}_2^{[\bar{v}(t'-t),a]}(x,x')$  only the usual
  logarithmic sum 
  \begin{align}
    \bar{R}_{-1,0}^{[s,a]}
    &=
      \sum_{q>0}
      \frac{\EXP^{\IMI qs}\EXP^{-aq}}{n_q}
    \\&
    =
    -\ln\Big(1-\EXP^{\frac{2\pi}{L}(\IMI s-a)}\Big)
    \stackrel{L\to\infty}{\longrightarrow}
    -\ln\Big(\frac{2\pi}{L}(a-\IMI s)\Big)
    \nonumber
  \end{align}
  survives, so that the contributions to the Green function for $L$
  $\to$ $\infty$ become
  \begin{align}
    M(\tau,\nu,x,t,x',t')
    &=
      \bigg[\frac{a}{\IMI [x-x'-\bar{v}(t-t')]+a}\bigg]^{\nu^2}
      \,.
  \end{align}
  The Green function then takes the familiar power-law form
  \begin{align}
    \label{eq:G-result-translinv}
    &\Ggtr_{\eta}(x,x\PRIME;t)
      =M^{>}_{x,x',t}
    \\&~~~~\times
    \bigg[\frac{-\IMI a}{x-x'-\bar{v}t-\IMI a}\bigg]^{1+v^2}
    \bigg[\frac{\IMI a}{x-x'+\bar{v}t+\IMI a}\bigg]^{v^2}
    \,,\nonumber
  \end{align}
  with 
  dependence on only $x-x'\pm\IMI\bar{v}t$. The  
  interaction-dependent  exponent, $v^2$ $=$
  $(\sqrt{v_{\calN}/v_{\calJ}}$ $-$
  $\sqrt{v_{\calJ}/v_{\calN}})^2/4$, depends only on the velocity
  ratio of $v_{\calN}/v_{\calJ}$, which is a characteristic
  feature of the Luttinger liquid that remains valid even for weakly
  nonlinear dispersions.\cite{haldane_luttinger_1981} Furthermore, in the
  translationally invariant case without interaction we have $\gamma$ $=$
  $0$ and hence $v$ $=$ $0$, so that only the first factor with unit
  exponent correctly remains in~\eqref{eq:G-result-translinv}.

  \subsubsection{Weak quadratic position dependence of the interactions}

  Next we consider position-dependent potentials that are regular at
  the origin, i.e., $V(x)$ $=$ $V(0)$ $+$ $V''(0)x^2/2$ $+$ $O(x^4)$,
  which is sketched in Fig.~\ref{fig:sketch}a for the repulsive case.
  From~\eqref{eq:xi+r0} we find for the function $r_0(x)$ that
  \begin{align}
    r_0(x)
    &=
      r_0'(0)\,x+\frac16r_0'''(0)\,x^3 + O(x^5)
      \,,
    \\
    r_0'(0)
    &=
      \frac{2\pi\bar{v}}{\bar{\gamma}(2\pi\VF+V(0))}\equiv\alpha
      \,,~
      r_0''(0)
      =
      0
      \,,\nonumber
    \\
    r_0'''(0)
    &=
      \frac{-2\pi\bar{v}V''(0)}{\bar{\gamma}(2\pi\VF+V(0))^2}\equiv6\beta
      \,.\nonumber
  \end{align}
  We will be interested in the asymptotic behavior of Green
  functions (rather than their periodicity in $L$) and thus will
  eventually take the limit $L$ $\to$ $\infty$. We therefore
  consider a weak correction to the linear behavior $r'(0)$, i.e.,
  \begin{align}
    r_0(x)
    &=
      \alpha x
      +
      \beta x^3 + O(\beta^2 x^5)\,,
    &
      \beta
    &=
      \frac{\text{const}}{L^2}
      \,,\label{eq:cubic-r0-approx-withbeta}
    \\
    x_0(r)
    &=
      \bar{\alpha} r
      -
      \bar{\beta} r^3 + O(\bar{\beta}^2 r^5)\,,
    &
      \bar{\alpha}
    &=
      \frac{1}{\alpha}
      \,,
      ~
      \bar{\beta}
      =
      \frac{\beta}{\alpha^4}
      \,.\nonumber
  \end{align}
  For the potential this means
  \begin{align}
    \!\!\!\!
    V(x)
    &
      =
      V(0)-\frac{6\pi\tilde{v}_{\text{F}}\beta}{\alpha}x^2
      +
      O\Big(\frac{x^4V_0}{L^4}\Big)
      \,.\!\!\!\!
  \end{align}
  The following choice of coefficients $\bar{R}_q$ turn out to produce this
  behavior,
  \begin{align}
    \bar{R}_q
    &=
      e^{-\REXPO|q|L/\pi}
      \,,\label{eq:Rq-exponential}
  \end{align}
  where $\REXPO$ is positive dimensionless
  parameter, because from~\eqref{eq:x0-from-Rq} we find
  \begin{align}
    x_0(r)
    &=
      r+
      \frac{L}{\pi}
      \arctan\frac{\sin\frac{2\pi r}{L}}{\EXP^{2\REXPO}-\cos\frac{2\pi r}{L}}
      \,,
  \end{align}
  which for small $|x/L|$ corresponds
  to~\eqref{eq:cubic-r0-approx-withbeta} with
  \begin{subequations}%
    \begin{align}%
      \bar{\alpha}
      &=
        \cth\REXPO
        \,,
      &
        \bar{\beta}
      &=
        \frac{\pi^2}{3}
        \frac{\ch\REXPO}{\sh^3\REXPO}
        \Big(\frac{2}{L}\Big)^2
        \,,
      \\
      \alpha
      &=
        \tnh\REXPO
        \,,
      &
        \beta
      &=
        \frac{\pi^2}{3}
        \frac{\sh\REXPO}{\ch^3\REXPO}
        \Big(\frac{2}{L}\Big)^2
        \,.
    \end{align}%
  \end{subequations}
  The functions~\eqref{eq:Rfunc-def} are evaluated from~\eqref{eq:Rq-exponential} as 
  \begin{subequations}
    \begin{align}
      \bar{R}_{-1,n}^{[s,a]}
      &=
        -\ln\Big(1-\EXP^{\frac{2\pi}{L}(\IMI s-a-n\REXPO L/\pi)}\Big)
        \,,
      \\
      \bar{R}_{m,n}^{[s,a]}
      &=
        \frac{\EXP^{(-\frac{2\pi}{L}(\IMI s-a)+2n\REXPO)m}}{[\EXP^{-\frac{2\pi}{L}(\IMI s-a)+2n\REXPO}-1]^{m+1}}
        \,.~~(m=0,1)
    \end{align}%
  \end{subequations}
  For large $L$, the last logarithmic term in the exponent of~\eqref{eq:Msigma-result}
  then dominates, containing
  \begin{align}
    \bar{S}_2^{[s,a]}(x,y)
    &=
      -\ln\bigg[
      \frac{
      \sh\frac{\pi}{L}(\IMI[s+r_0(x)-r_0(y)]-a)
      }{
      \sh\frac{\pi}{L}(\IMI[s+r_0(x)]-a-\REXPO L/\pi)
      }\nonumber
    \\&~~~~\times
    \frac{
    \sh\frac{\pi}{L}(\IMI s-a-\REXPO L/\pi)
    }{
    \sh\frac{\pi}{L}(\IMI[s-r_0(y)]-a-\REXPO L/\pi)
    }
    \bigg]
    \,.
  \end{align}
  To leading order in $x/L$, $x'/L$, the Green function then becomes
  \begin{align}
    &\Ggtr_{\eta}(x,x\PRIME;t)
      =
      M^{>}_{x,x',t}
      \label{eq:G-result-weakpositiondependence}
    \\&\times
    \bigg[\frac{-\IMI a}{\alpha(x-x')-\bar{v}t-\IMI a}\bigg]^{1+v^2}
    \bigg[\frac{\IMI a}{\alpha(x-x')+\bar{v}t+\IMI a}\bigg]^{v^2}
    \,,\nonumber
  \end{align}
  i.e., translational invariance is only broken in finite-size
  corrections.

  Note that according
  to~\eqref{eq:G-result-weakpositiondependence} a fermionic
  single-particle perturbation near $x$ $=$ $0$, as measured by the
  Green function, propagates with velocity $\bar{v}/\alpha$ $=$
  $\bar{v}/r_0'(0)$ $=$ $\bar{\gamma}(\VF+V(0)/(2\pi)$. This which differs
  from the translationally invariant
  case~\eqref{eq:luttingerliquid-vrelation} with corresponding
  velocity $\bar{\gamma}(\VF+V_0/(2\pi)$ for which only the
  position-averaged interaction $V_0$ matters. For the Luttinger
  droplet, the position dependence of $V(x)$ is thus observable in the
  propagation velocity described by the Green function. This can be
  observed in more detail for a stronger position dependence of
  $V(x)$, as discussed in the next subsection.

  We also note that the exponent $v^2$ (expressed in terms of $\gamma$
  in~\eqref{eq:ham_par_bos}) is no longer related only to the velocity
  ratio of $v_{\calN}/v_{\calJ}$, hence this feature of the Luttinger
  liquid is also no longer present.

  \subsubsection{Piecewise constant interaction potential}\label{subsubsec:piecewise}

  As a minimal example which explicitly breaks the translational
  invariance of the Green function, we consider an interaction
  potential that is piecewise constant,
  \begin{align}
    \label{eq:piecewiseV}
    V(x)
    &=
      \begin{cases}
        V(0)&\text{~if~}|x|<R\,,
        \\
        V(\tfrac{L}{2})&\text{~if~}|x|>R\,,
      \end{cases}                          
  \end{align}
  i.e., the particles interact differently inside a central region and
  outside of it, as depicted in Fig.~\ref{fig:sketch}b for the
  repulsive case. The average of this function is given by
  \begin{align}
    V_0
    &=
      r\,V(0)+
      (1-r)\,V(\tfrac{L}{2})
      \,,~~
      r=\frac{2R}{L}
      \,.
  \end{align}
  Here $r$ is the fraction of the central region with interaction
  $V(0)$, which tends to zero if we consider a fixed finite central
  interval of width $2R$ but let $L$ tend to infinity, see below.
  For the potential~\eqref{eq:piecewiseV} we find
  \begin{subequations}%
    \begin{align}%
      \bar{v}
      &=
        \frac{\bar{\gamma}}{2\pi}\frac{1}{rs+(1-r)\tilde{s}}
        \,,
      \\
      r_0(x)
      &=
        \begin{cases}
          \alpha x
          &\text{~if~}|x|\leq R\,,
          \\
          \tilde{\alpha}x+\text{sgn}(x)(\alpha-\tilde{\alpha})R
          &\text{~if~}|x|\geq R\,,
        \end{cases}
            \label{eq:r0-result-box}
      \\
      \bar{R}_q
      &=
        r
        \,
        \frac{V(0)-V(\tfrac{L}{2})}{2\pi\VF+V(0)}
        \,
        \frac{\sin n_q\pi r \alpha}{n_q\pi r \alpha}
        \,,~~(q>0)
        \,,\label{eq:Rq-result-box}
    \end{align}%
  \end{subequations}%
  with the abbreviations
  \begin{align}%
    s
    &=
      \frac{1}{2\pi\VF+V(0)}
      \,,&
           \tilde{s}
    &=
      \frac{1}{2\pi\VF+V(\tfrac{L}{2})}
      \,,
    \\
    \alpha
    &=
      r_0'(0)=\frac{s}{\tilde{s}+(s-\tilde{s})r}
      \,,&
           \tilde{\alpha}
    &=
      r_0'(\tfrac{L}{2})=\frac{\tilde{s}}{\tilde{s}+(s-\tilde{s})r}
      \,.\nonumber
  \end{align}%
  From now on we consider only fixed finite $R$ and let $L$ $\to$
  $\infty$, i.e., $r$ $\to$ $0$. The second fraction
  in~\eqref{eq:Rq-result-box} involving the sine function can then
  be replaced by unity. In this limit the
  summations~\eqref{eq:Rfunc-def} evaluate to
  \begin{align}
    \bar{R}_{m,n}^{[s,a]}
    &=\Bigg[r
      \,
      \frac{V(0)-V(\tfrac{L}{2})}{2\pi\VF+V(0)}
      \Bigg]^n
      \bar{R}_{m,0}^{[s,a]}
      \,.
  \end{align}
  The logarithmic term in $\bar{S}_2^{[s,a]}(x,y)$ then again provides 
  the leading term in~\eqref{eq:Msigma-result} for $L$ $\to$ $\infty$,
  \begin{align}
    &M(\tau,\nu,x,t,x',t')\nonumber
    \\&~~=
    \bigg[\frac{a}{\IMI [r_0(x)-r_0(x')-\bar{v}(t-t')]+a}\bigg]^{\nu^2}
    \,.
  \end{align}
  The Green function then takes a power-law form with piecewise
  linear argument
  \begin{align}
    \label{eq:resultG_piecewise}
    \Ggtr_{\eta}(x,x\PRIME;t)
    &=M^{>}_{x,x',t}
    \\\nonumber&~~~~\times
                 \bigg[\frac{-\IMI a}{r_0(x)-r_0(x')-\bar{v}t-\IMI a}\bigg]^{1+v^2}
    \\\nonumber&~~~~\times
                 \bigg[\frac{\IMI a}{r_0(x)-r_0(x')+\bar{v}t+\IMI a}\bigg]^{v^2}
                 \!,
  \end{align}
  with the exponent $v^2$ given in terms of $\gamma$
  in~\eqref{eq:ham_par_bos}.  As listed in~\eqref{eq:r0-result-box},
  in the present case $r_0(x)$ is piecewise
  linear in $x$ with a change in slope at $|x|$ $=$ $R$. Hence if $x$
  and $x'$ lie both inside or both outside the central region, the
  Green function is essentially the same as in the case of weak
  position dependence~\eqref{eq:G-result-weakpositiondependence} or
  the translationally invariant case~\eqref{eq:G-result-translinv},
  respectively. However, if only one of $x$ and $x'$ is inside the
  central region, the two coordinates enter with different prefactors
  into the Green function, breaking its translational invariance.  The
  Green function~\eqref{eq:resultG_piecewise} and velocity
  relation~\eqref{eq:luttingerdroplet-vrelation} indicate that for the
  interaction potential~\eqref{eq:piecewiseV} the Luttinger
  droplet~\eqref{eq:dropletHfermionicLR-intro} is 
  distinguishable from the Luttinger liquid.
  
  Moreover, the Green function~\eqref{eq:resultG_piecewise} shows that
  a fermionic single-particle perturbation created at position $x$
  will initially propagate with velocity $\bar{v}/r_0'(x)$ $=$
  $\bar{\gamma}(\VF+V(x)/(2\pi)$, which is piecewise constant in the
  present case. As might have been expected, the position dependence
  of $V(x)$ thus translates into a position-dependent `local'
  propagation velocity. Its relation to the other excitation
  velocitues of the Luttinger droplet model will be discussed the next
  section.
  
  \section{Towards a Luttinger droplet paradigm}\label{sec:dropletparadigm}

  The translationally invariant Tomonaga-Luttinger model obeys the
  relations~\eqref{eq:translationallyinvariantcase} between excitation
  velocities and Green function exponents, i.e., in our notation
  between $\bar{v}$, $v_{\calN}$, $v_{\calJ}$, and $\gamma$. In
  particular, the dressed Fermi velocity $\bar{v}$ appears in the
  Green function~\eqref{eq:G-result-translinv} as the velocity with
  which a fermion $\PSIadd{\eta}{x}$ propagates when added to the
  Luttinger liquid ground state.
  For the Luttinger droplet model~\eqref{eq:dropletHfermionicLR-intro}
  (with linear dispersion) we found different relations between the
  excitation velocities and $\gamma$, as given
  in~\eqref{eq:luttingerdroplet-vrelation}.  Furthermore, the Green
  functions of Sec.~\ref{sub:dropletgreen} show that a fermion
  $\PSIadd{\eta}{x}$, inserted into the Luttinger droplet ground state
  at position $x$, initially propagates with velocity
  $\bar{v}/r_0'(x)$.  This behavior was observed explicitly for a weak
  and piecewise constant position dependence of the interaction
  potential $V(x)$ in~\eqref{eq:G-result-weakpositiondependence} and
  \eqref{eq:resultG_piecewise}, respectively.  It can be traced
  to~\eqref{eq:xi+r0}, where a phase $r_0(x)k$ appears in the exponent
  of the eigenfunctions $\xi_k(x)$ of the refermionized
  model~\eqref{eq:refermionization:sigma}. We can therefore expect that
  a `local' propagation velocity of fermionic perturbations,
  \begin{align}
    v^{\text{loc}}(x)
    &=
      \frac{\bar{v}}{r_0'(x)}
      =
      \bar{\gamma}\bigg(\VF+\frac{V(x)}{2\pi}\bigg)
      \,,
  \end{align}
  will appear in the Green function also for more general $V(x)$.
  Compared to the translationally invariant case this is a new range
  of velocities, which we will now relate to the
  other excitation velocities of the Luttinger droplet.
  
  For this purpose we first seek to characterize the scales of
  $v^{\text{loc}}(x)$. One way to do this uses its arithmetic and
  harmonic averages over the entire system. For these we find
  \begin{subequations}%
    \begin{align}%
      \!\!
      \overline{v^{\text{loc}}}
      \equiv
      \langle\langle v^{\text{loc}}(x) \rangle\rangle_{\text{arith}}
      &\equiv
        \hphantom{\bigg[}\,
        \int\!\frac{\DX}{L}\,v^{\text{loc}}(x)\,
        \hphantom{\bigg]^{-1}}
      =
      \bar{\gamma}\tilde{v}_{\text{F}}
      \,,\!\!
      \\
      \!\!
      \langle\langle v^{\text{loc}}(x) \rangle\rangle_{\text{harm}}
      &\equiv
        \bigg[\int\!\frac{\DX}{L}\,\frac{1}{v^{\text{loc}}(x)}\bigg]^{-1}
      =
      \bar{v}
      \,,
    \end{align}%
    \label{eq:dropletaveragecelocities}%
  \end{subequations}%
  where, as above, $\tilde{v}_{\text{F}}$ $=$ $\VF$ $+$
  $V_0/(2\pi)$. For general $V(x)$ these two averages are different,
  but coincide in the translationally invariant case.  With the
  excitations of the Luttinger droplet characterized by the velocities
  $\bar{v}$, $\overline{v^{\text{loc}}}$, $v_{{\calN}}$,
  $v_{{\calJ}}$, we then obtain their interrelation
  from~\eqref{eq:luttingerdroplet-vrelation},
  \begin{subequations}%
    \label{eq:dropletrelations}
    \begin{align}%
      \bar{v}
      &=
        c_{{\calN}}(\gamma)\;v_{{\calN}}
        +
        c_{{\calJ}}(\gamma)\;v_{{\calJ}}
        \,,
      \\
      \overline{v^{\text{loc}}}
      &=
        c^{\text{loc}}_{{\calN}}(\gamma)\,v_{{\calN}}
        +
        c^{\text{loc}}_{{\calJ}}(\gamma)\,v_{{\calJ}}
        \,,
    \end{align}%
  \end{subequations}%
  where the prefactors are given by
  \begin{subequations}%
    \label{eq:coefficients}%
    \begin{align}%
      c_{{\calN},{\calJ}}(\gamma)
      &=
        \frac{(\gamma-\bar{\gamma}\gamma_3)\mp(1-\bar{\gamma})}{2(\gamma-\gamma_3)}
        \,,
      \\
      c^{\text{loc}}_{{\calN},{\calJ}}(\gamma)
      &=
        \bar{\gamma}\,
        \frac{(\pm1-\gamma_3)}{2(\gamma-\gamma_3)}
        \,,
    \end{align}%
  \end{subequations}%
  Furthermore $\gamma$, which characterizes the relative strength of
  interbranch interactions, determines the Green function exponent
  $v^2$ according to~\eqref{eq:ham_par_bos}. 
  The dependence of the
  coefficients~\eqref{eq:coefficients} on $\gamma$ is shown in
  \begin{figure}[tb]
    \centering
    \includegraphics[width=\columnwidth]{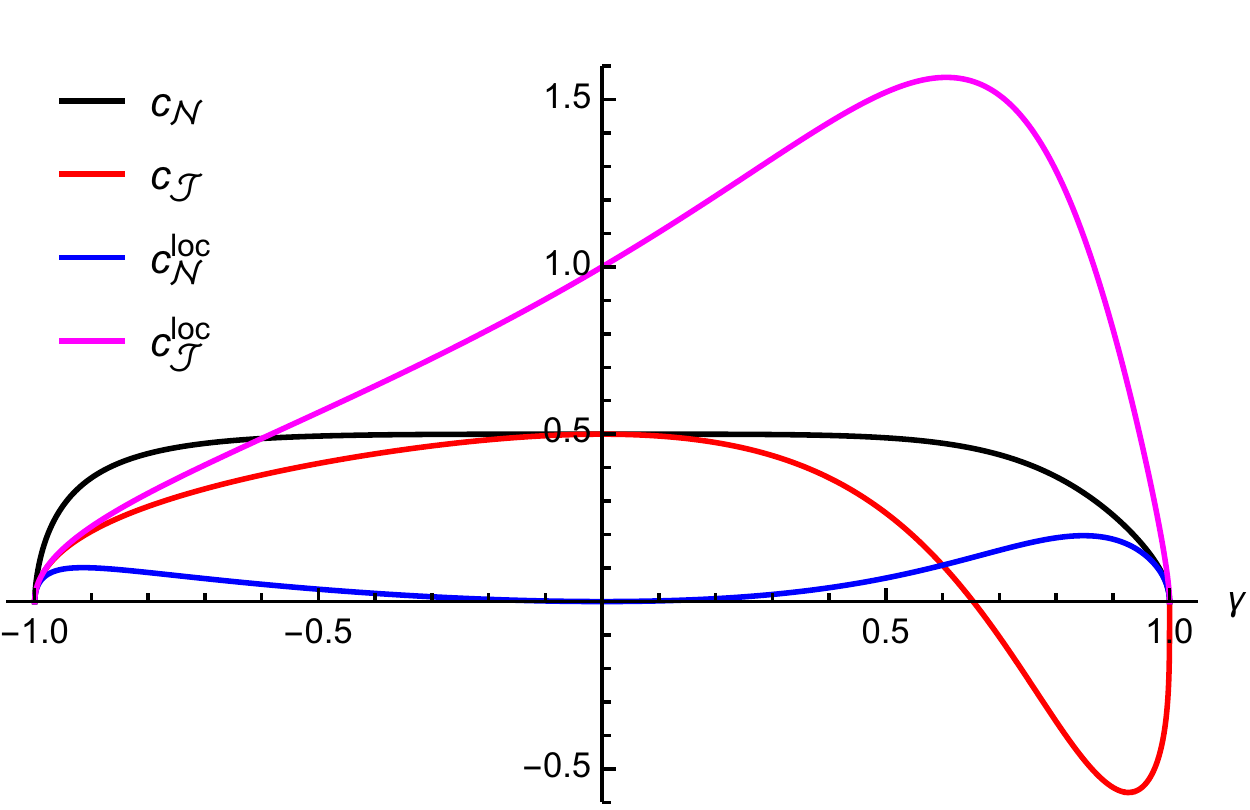}
    \caption{Coefficients~\eqref{eq:coefficients} in the linear
      relation~\eqref{eq:dropletrelations} between excitation
      velocities in the Luttinger droplet
      model~~\eqref{eq:dropletHfermionicLR-intro} as a function of the
      interaction parameter $\gamma$ given by
      \eqref{eq:gamma-specialcase}, \eqref{eq:luttingerdroplet-gamma}.\label{fig:coeffs}}
  \end{figure}
  Fig.~\ref{fig:coeffs}. We note that for $\gamma$ $=$ $0$, the two
  branches in the Hamiltonian do not mix; in this case $v_{{\calN}}$
  and $v_{{\calJ}}$ contribute equally to $\bar{v}$ and
  $\overline{v^{\text{loc}}}$ equals $v_{{\calN}}$. On the other hand,
  for only interbranch interactions ($\gamma$ $\to$ $\pm1$),
  $\bar{\gamma}$ vanishes and hence so do $v_{{\calN},{\calJ}}$.

  A preliminary physical interpretation of the
  velocities~\eqref{eq:dropletaveragecelocities} might be that
  $\bar{v}$ plays the role of group velocity, as $\bar{v}q$ is the
  energy of a bosonic excitation
  in~\eqref{eq:luttingerdroplet-diagonalized} which involves a
  nonlocal and mixed-flavor superposition of original fermions. On the
  other hand, since $v^{\text{loc}}(x)$ plays the role of a local
  phase velocity, its scale is presumably captured by the arithmetic
  average $\overline{v^{\text{loc}}}$.  Note that for the
  translationally invariant case $\bar{v}$ $=$
  $\bar{\gamma}\tilde{v}_{\text{F}}$, and indeed the group velocity
  and (position-independent) phase velocity are both given by
  $\bar{v}$, cf.~\eqref{eq:luttingerdroplet-diagonalized},
  \eqref{eq:translationallyinvariantcase},
  \eqref{eq:G-result-translinv}.
  
  We conclude that for the Luttinger droplet
  model~\eqref{eq:dropletHfermionicLR-intro} the quantities $\bar{v}$,
  $\overline{v^{\text{loc}}}$, $v_{{\calN}}$, $v_{{\calJ}}$, and
  $\gamma$ are related, extending the Luttinger liquid relations
  between $\bar{v}$, $v_{\calN}$, $v_{\calJ}$, $\gamma$ to the
  position-dependent case. However, it remains to clarify how the
  relations~\eqref{eq:dropletrelations} evolve away from the special
  case~\eqref{eq:dropletHfermionicLR-intro-parameters}.  Furthermore,
  in order to be regarded as a paradigm for one-dimensional electronic
  systems with position-dependent interactions, these relations would
  have to remain valid also for weak nonlinearites in the
  dispersion. Both of these questions would therefore be worthwhile to
  address, e.g., by perturbative methods.


  \section{Conclusion}\label{sec:conclusion}

  Using higher-order bosonization identities, i.e., Kronig-type
  relations with finite momentum transfer, we solved the Luttinger
  droplet model~\eqref{eq:dropletHfermionicLR-intro} for a large class
  of position-dependent interactions and arbitrary one-particle
  potentials. While the diagonalized Hamiltonian has the same operator
  expression as for the Luttinger liquid, the relation between its
  velocity parameters is not fulfilled in general, as the bosonic
  excitations and particle number changes involve different averages
  of the interaction potential over all positions. Similarly the Green
  functions retain their power-law form for weak position-dependence
  of the interaction potential, but their exponents also no longer
  depend only on the ratio of excitation velocities for
  particle-number changes. For weak position-dependent interactions
  the Luttinger-liquid characteristics are rather robust regarding
  their functional form, although the interrelation of the dressed
  scales and exponents is somewhat different. On the other hand, for
  an interaction potential with different (e.g., constant) values
  inside or outside a central region of finite width, not only are the
  Luttinger-liquid velocity relations modified, but also the Green
  function is no longer translationally invariant and exhibits a
  position-dependent propagation velocity of single-particle
  excitations. This may mean that the group velocity of such an
  excitation differs from its (position-dependent) phase velocity, in
  contrast to the Luttinger liquid.  We conclude that the
  Luttinger droplet model has a ground state with different
  characteristics than the Luttinger liquid. It remains to be seen how
  the velocity relations obtained
  for~\eqref{eq:dropletHfermionicLR-intro} evolve for more general
  one-dimensional models with position-dependent interactions, and
  whether a Luttinger droplet paradigm emerges for them.

  \acknowledgments
  
  The authors would like to thank Matthias Punk and Jan von Delft for
  valuable discussions. M.K. would also like to thank Sebastian Diehl,
  Erik Koch, Volker Meden, Lisa Markhof, Aditi Mitra, Herbert Schoeller, and Eva
  Pavarini for useful discussions. S.H. gratefully acknowledges
  support by the German Excellence Cluster Nanosystems Initiative
  Munich~(NIM) and by the Deutsche Forschungsgemeinschaft under
  Germany's Excellence Strategy EXC-2111-390814868.  M.K. was
  supported in part by Deut\-sche For\-schungs\-ge\-mein\-schaft under
  Pro\-jekt\-num\-mer 107745057 (TRR 80) and performed part of this
  work at the Aspen Center for Physics, which is supported by National
  Science Foundation grant~PHY-1607611.
 
  \bibliographystyle{apsrev4-1}
  \bibliography{bosopapbib_short}

\end{document}